\documentclass[letterpaper,hyper]{JHEP3}
\usepackage[centertags]{amsmath}
\usepackage{url}
\usepackage{wrapfig}
\usepackage{array,multirow}
\usepackage{amssymb}
\usepackage{graphicx}
\usepackage{color}
\usepackage{mathtools}
\usepackage{wasysym}
\usepackage[punctsep]{collref}
\collectsep[]{;}

\usepackage{epsfig}

\def\l{\lambda}
\def\la{\lambda}
\def\lad{\hat{\lambda}}
\def\lbd{\hat{\bar{\lambda}}}

\def\lb{\bar{\lambda}}

\def\mb{\bar{\mu}}

\def\ep{\epsilon}

\def\Or[#1]{{\text{O}}\left({#1}\right)}
\def\dotl[#1,#2]{\left\langle #1,\, #2 \right\rangle}
\def\dotlb[#1,#2]{\left\langle #1,\, #2 \right\rangle}
\def\dotlm[#1,#2]{\left[ #1,\, #2 \right]}
\def\dotp[#1,#2]{(\vect{#1} \cdot\vect{#2})}
\def\aff[#1,#2]{\hat{#1}(#2)}
\def\n4sym{{\cal N}=4 SYM}
\def\>{\rangle}
\def\<{\langle}
\def\weight[#1,#2,#3]{\{(#1),#2,#3\}}
\def\ads[#1]{$\text{AdS}_{#1}$}
\def\lrdel{\overset{\leftrightarrow}{\nabla}}
\newcommand{\be}{\begin{equation}}
\newcommand{\ee}{\end{equation}}
\newcommand{\ba}{\begin{align}}
\newcommand{\ea}{\end{align}}
\newcommand{\bs}{\begin{split}}
\def\sess\end{split}
\newcommand{\vect}[1]{{\boldsymbol{#1}}}
\newcommand{\norm}[1]{|{\boldsymbol{#1}}|}
\title{Four Point Functions of the Stress Tensor and Conserved Currents in AdS$_4$/CFT$_3$}
\author{Suvrat Raju \\ 
Harish-Chandra Research Institute, \\ Chatnag Road, Jhunsi, \\ Allahabad 211019.}
\abstract{
We compute four point functions of the stress tensor and conserved currents in AdS$_4$/CFT$_3$  using bulk perturbation theory. We work at treel level in the bulk theory, which we take to be either pure gravity or Yang Mills theory in AdS.  We bypass the tedious evaluation of Witten diagrams using recently developed recursion relations for these correlators. In this approach, the four point function is obtained as the sum of residues of a rational function at easily identifiable poles. We write down an explicit formula for the four point correlator with arbitrary external helicities and momenta. We verify that, precisely as conjectured in a companion paper, the Maximally Helicity Violating (MHV) amplitude of gravitons or gluons appears as the coefficient of a specified singularity in the MHV stress-tensor or current correlator.  We comment on the remarkably simple
analytic structure of our answers in momentum space. 
}
\preprint{HRI/ST/1202}
\begin{document}
\section{Introduction}
It is remarkable that even fifteen years after AdS/CFT was discovered \cite{Maldacena:1997re} there are very few explicit computations of boundary correlation functions, at four points or higher, from the bulk point of view \cite{Liu:1998ty,Freedman:1998tz,Arutyunov:2000py,Bartels:2009jb}. In principle, such computations are straightforward: we need to link interaction vertices with bulk-bulk and bulk-boundary propagators, and integrate over their positions \cite{Witten:1998qj,Gubser:1998bc}. However, in practice these computations are difficult for two reasons. First, it is hard to do the bulk integrals in closed form, although this can be sidestepped by using clever tricks \cite{D'Hoker:1999ni} or by transforming to Mellin space \cite{Fitzpatrick:2011ia,Penedones:2010ue,Paulos:2011ie,Mack:2009mi,Mack:2009gy,Costa:2011dw,Costa:2011mg}. However, in theories like gravity the interaction vertices themselves are very complicated. For example, the four point vertex, even in four dimensional flat space, contains 2,850 terms \cite{DeWitt:1967uc} and is even more complicated in AdS. Consequently, the four point function of the stress tensor has never before been computed explicitly.

In this paper, we point out that going to momentum space on the boundary in AdS$_4$ solves both these problems at once for correlators of conserved currents of the stress tensor. In AdS$_4$, the momentum-space bulk to boundary and bulk to bulk propagators for gluons and gravitons can be written in terms of elementary functions and so doing $z$-integrals is very simple. 

Furthermore, by generalizing insights from flat space computations of gravity and Yang-Mills amplitudes \cite{Witten:2003nn,Britto:2004ap,Britto:2005fq,ArkaniHamed:2008gz,ArkaniHamed:2009dn,Forde:2007mi, Spradlin:2008bu, Raju:2009yx,Lal:2009gn,Lal:2010qq, Boels:2010nw, Mason:2008jy,Nguyen:2009jk}, we are able to sidestep the tedious evaluation of interaction vertices. Instead, we compute explicit expressions for the four point functions of the stress tensor by using the known three point functions as input and combining this with a knowledge of the analytic properties of the correlators.

More specifically, we implement the recursion relations proposed in a companion paper \cite{Raju:2012zr}. These relations allow us to write down a formula for the four point function in terms of residues of a rational function, at specific poles. This rational function is obtained through the product of two deformed three point ``transition amplitudes'' as we explain in more detail in section \ref{secformula}. 

In \cite{Raju:2012zr}, it was pointed out that momentum space also allows
us to take an elegant ``flat space limit.'' Here we take the flat space limit of our results for correlation functions of the stress tensor
and obtain exactly the famous formulas for maximally helicity violating (MHV) amplitudes of gluons and gravitons in 4 dimensional flat space. 

To facilitate this comparison, we write our results for MHV correlators in the spinor helicity formalism that was originally
developed for four dimensional flat space amplitudes but, as pointed out
in \cite{Maldacena:2011nz}, is also useful for computations in AdS$_4$. 

Apart from this flat space limit, our answers have some other interesting
structural properties. For example, we can immediately see the contribution of the entire {\em conformal block} of the stress tensor itself in the correlator. Once again momentum space makes this simple. Here, 
we just have to multiply correlators to get the contribution of the primary and all its descendants rather than worrying about the complicated
expressions for conformal blocks in position space. 

Second, the transition amplitudes that appear in our computations
are finite. So we avoid the divergences that appear in momentum space AdS integrals from the region near the boundary. An interesting consequence of this is that the correlators can be written as {\em rational functions} of the external spinors, and norms of partial sums of the external momenta. In particular we do not find any logarithms in our answers. We comment more on this interesting fact
in the discussion section

A brief overview of this paper is as follows. We start by reviewing the
spinor helicity formalism in section \ref{secspinoreview}. The four point computations require three point transition amplitudes, which are very similar to
correlators, but are obtained by replacing a bulk to boundary propagator
with a normalizable mode. We compute these three point functions for
Yang-Mills theory and gravity in section \ref{secthreept}. In section \ref{secformula}, we use these results to write down an explicit formula for the four point function in terms of residues
of a rational function at pre-specified poles. In section \ref{secmhvym}, we evaluate this formula for correlators of conserved currents, with two positive helicity and two negative helicity insertions, using the spinor helicity formalism and verify that, in the flat space limit, it reduces to the 
scattering amplitude of gluons, as conjectured in \cite{Raju:2012zr}. In section \ref{secmhvgrav}, we evaluate this formula for stress tensor correlators, with the same combination of helicities, and, once again verify, that in the flat space limit it reduces to the maximally helicity violating amplitude of gravitons. 

The main idea of this paper is presented in \ref{secformula} and the reader who is not interested in the fine details of four point correlators can skip straight to this section. Moreover, we should warn the reader that some of the computations in section \ref{secmhvym} and \ref{secmhvgrav} are a little heavy on algebra. For this reason, we have provided a Mathematica program (available from the source of the arXiv version of this paper) that can be used to automate the formulas that are implemented there.

\section{Spinor Helicity Formalism \label{secspinoreview}}
We start by reviewing the spinor helicity formalism for correlation 
functions in 3 dimensional conformal field theories that was introduced
in \cite{Maldacena:2011nz}. 

We will use the mostly positive metric. So, for two vectors on the boundary
\begin{equation}
\vect{k} \cdot \vect{k} = (k_1)^2 + (k_2)^2 - (k_0)^2.
\end{equation}
In this paper, just as in \cite{Raju:2012zr} we use bold-face for vectors but not their components. We use  $i,j$ etc. for boundary spacetime indices and $\mu, \nu$ etc. for bulk spacetime indices. We use $m, n$ etc. to index particle-number but one difference from \cite{Raju:2012zr} is that here it is convenient to place this index in subscripts rather than superscripts. Also, the components of a momentum vector come with a naturally lowered index. 

Our $\sigma$ matrix conventions are the following
\begin{equation}
\begin{split}
\sigma^0_{\alpha \dot{\alpha}} = \begin{pmatrix}1&0\\0&1\end{pmatrix}, \quad \sigma^1_{\alpha \dot{\alpha}} = \begin{pmatrix}0&1\\1&0\end{pmatrix}, \\
\sigma^2_{\alpha \dot{\alpha}} = \begin{pmatrix}0&-i\\i&0\end{pmatrix}, \quad \sigma^3_{\alpha \dot{\alpha}} = \begin{pmatrix}1&0\\0&-1\end{pmatrix}.
\end{split}
\end{equation}

Given a three momentum $\vect{k} = (k_0, k_1, k_2)$, we convert it into 
spinors using
\begin{equation}
k_{\alpha \dot{\alpha}} = k_0 \sigma^0_{\alpha \dot{\alpha}} +  k_1 \sigma^1 +  k_2 \sigma^2 +  i |\vect{k}| \sigma^3 = \la_{\alpha} \lb_{\dot{\alpha}},
\end{equation}
where
\begin{equation}
|\vect{k}| \equiv \sqrt{\vect{k} \cdot \vect{k}} = \sqrt{k_1^2 + k_2^2 - k_0^2}.
\end{equation}
If $\vect{k}$ is spacelike to start with, then the $\sigma^3$ component will be imaginary. 

In components, we have the following expressions for the spinors
\begin{equation}
\begin{split}
\la &= \bigl(\sqrt{k_0 + i |\vect{k}|}, {k_1 + i k_2 \over \sqrt{k_0 + i |\vect{k}|}} \bigr), \\
\lb &= \bigl(\sqrt{k_0 + i |\vect{k}|}, {k_1 - i k_2 \over \sqrt{k_0 + i |\vect{k}|}} \bigr).
\end{split}
\end{equation}

We have the freedom to rescale the spinors by any complex number: $\la \rightarrow \alpha \la,~\lb \rightarrow {1 \over \alpha} \lb$ without changing the momentum.  If we do this with spinors corresponding to an external particle, then this 
rescales the polarization vectors and amplitudes pick up a simple phase. However, when we use the recursion relations, we also need spinors for an internal particle when we cut a propagator to form the product of two amplitudes. In such cases,  $\la_{\rm int}$ and $\lb_{\rm int}$ always come together and so we can rescale them without any physical effect at all. For example, we could choose:
\begin{equation}
\label{lint}
\begin{split}
\la_{\rm int} &= \bigl(1, {k_1 + i k_2 \over k_0 + i |\vect{k}|} \bigr), \\
\lb_{\rm int} &= \bigl(k_0 + i |\vect{k}|, k_1 - i k_2 \bigr).
\end{split}
\end{equation}

We can raise and lower spinor indices using the $\epsilon$ tensor. We choose the $\epsilon$
tensor to be $i \sigma_2$ for both the dotted and the undotted indices. This means that 
\begin{equation}
\epsilon^{0 1} = 1 = -\epsilon^{1 0},
\end{equation}
and spinor dot products are defined via
\be
\dotl[\la_1, \la_2] = \epsilon^{\alpha \beta} \la_{1 \alpha} \la_{2 \beta}, = \la_{1 \alpha} \la_{2}^{\alpha}, \quad \dotl[\lb_1, \lb_2] = \epsilon^{\dot{\alpha} \dot{\beta}} \lb_{1 \dot{\alpha}} \lb_{2 \dot{\beta}} = \la_{1 \dot{\alpha}} \lb_1^{\dot{\alpha}}. 
\ee

However, we should expect our expressions for CFT$_3$ correlators to only have a manifest $SO(2,1)$ invariance. This means that we might have mixed products between dotted and undotted indices. Such a mixed product extracts the $z$-component
of vector and is performed by
contracting with $\sigma^3$
\begin{equation}
\label{mixedproduct}
2 i |\vect{k}| = (\sigma^3)^{\alpha \dot{\alpha}} k_{\alpha \dot{\alpha}} \equiv  \dotlm[\la, \lb],
\end{equation}
The reader should note that we use square brackets only for this
mixed product; products of both left and right handed spinors are denoted by angular brackets. Second, we note that this mixed dot product is symmetric:
\begin{equation}
\dotlm[\la, \lb] = \dotlm[\lb, \la].
\end{equation}

When we take the dot products of two 3-momenta, we have
\begin{equation}
\begin{split}
&\vect{k} \cdot \vect{q} \equiv \bigl(k_1 q_1 + k_2 q_2 - k_0 q_0 \bigr) \\ &=  -{1 \over 2} \Big(\dotl[\la_k, \la_q] \dotlb[\lb_k, \lb_q] + {1\over 2} \dotlm[\la_k,\lb_k] \dotlm[\la_q,\lb_q] \Big).
\end{split}
\end{equation}
Note that we have made a choice of the metric that is mostly positive.

Another fact to keep in mind is that
\begin{equation}
\begin{split}
&\vect{k_1} + \vect{k_2} = \vect{k_3} \\ &\Rightarrow \la_1 \lb_1 + \la_2 \lb_2 = \la_3 \lb_3 + {1 \over 2} \bigl(\dotlm[\la_1,\lb_1] +  \dotlm[\la_2,\lb_2] - \dotlm[\la_3,\lb_3] \bigr) \sigma^3.
\end{split}
\end{equation}

We also need a way to convert dotted to undotted indices.  We write
\begin{equation}
\lad_{\dot{\alpha}} = \sigma^3_{\alpha \dot{\alpha}} \la^{\alpha}, \quad \lbd_{\alpha} = \sigma^3_{\alpha \dot{\alpha}} \lb^{\dot{\alpha}}.
\end{equation}
This has the property that
\begin{equation}
\dotlb[\mb, \lad] = \dotlm[\mb,\la],
\end{equation}
where the quantity on the right hand side is defined in \eqref{mixedproduct}.

With all this, we can write down polarization vectors for conserved currents. The polarization vectors for a momentum vector $\vect{k}$ associated with
spinors $\vect{\la}, \vect{\lb}$ are given by
\begin{equation}
\label{polarizationvects}
\begin{split}
&\epsilon^+_{\alpha \dot{\alpha}} = 2 {\lbd_{\alpha} \lb_{\dot{\alpha}} \over \dotlm[ \la, \lb]} =  {\lbd_{\alpha} \lb_{\dot{\alpha}} \over i \norm{k}}, \\ 
&\epsilon^-_{\alpha \dot{\alpha}} =  2 {\la_{\alpha} \lad_{\dot{\alpha}} \over \dotlm[\la, \lb]} = {\la_{\alpha} \lad_{\dot{\alpha}} \over i \norm{k}}.
\end{split}
\end{equation}
These vectors are normalized so that 
\begin{equation}
\label{normalizationpol}
\vect{\epsilon^+} \cdot \vect{\epsilon^{+}} = \vect{\epsilon^-} \cdot \vect{\epsilon^{-}} = 0, \quad \vect{\epsilon^+} \cdot \vect{\epsilon^{-}} = 2.
\end{equation}
Polarization tensors for the stress tensor are just outer-products of these vectors with themselves:
\be
e^{\pm}_{i j} = \ep^{\pm}_i \ep^{\pm}_j.
\ee

In this paper, we will compute three and four point functions functions of the stress tensor and conserved currents in momentum space:
\be
T(\vect{e_1}, \vect{k_1},\ldots,\vect{e_n}, \vect{k_n}) = e_{1 i_1 j_1} \ldots e_{n i_n j_n} \langle T^{i_1 j_1}(\vect{k^1}) \ldots T^{i_n j_n}(\vect{k^n}),
\ee
where 
\be
 \langle T^{i_1 j_1}(\vect{k^1}) \ldots T^{i_n j_n}(\vect{k^n}) \rangle \equiv  \int \langle {\cal T}\Big\{T^{i_1 j_1}(\vect{x_1}) \ldots T^{i_n j_n}(\vect{x_n}) \Big\} \rangle e^{i \sum_{m=1}^n\vect{k_m} \cdot \vect{x_m}} d^d x_m, 
\ee
and ${\cal T}$ is the time-ordering symbol. Given the explicit formulas
for polarization vectors, we can also label correlators using the helicity and momenta of the various insertions. For example:
\be
T(+,\vect{k_1},+, \vect{k_2},-, \vect{k_3}) \equiv T\big(\vect{e_1^+}, \vect{k_1}, \vect{e_2^+}, \vect{k_2},\vect{e_3^{-}}, \vect{k_3} \big).
\ee
We will use the same notation to refer to correlators of conserved currents, and the meaning should be clear from the context. 

\section{Three Point Transition Amplitudes \label{secthreept}}
In this section we compute three point transition amplitudes that are an essential building block for the four point computation. We start with Yang-Mills theory where the Feynman rules are quite easy to establish and then move on to gravity. 

We remind the reader that transition amplitudes are computed by replacing one bulk to boundary propagator with a normalizable mode. Below, we use spinors $\la_1, \ldots \la_3$, and $\lb_1, \ldots \lb_3$ to specify the three
momenta in the amplitude. We will use $p = -i \norm{k_3}$ to indicate that this
leg is distinguished because it is the one that is associated with the normalizable mode. 

\subsection{Yang Mills Theory}

Let us first review the form of the gauge-boson bulk to boundary propagator in AdS$_4$.  As we will see below for both gauge-bosons and gravitons
propagating in AdS$_4$, the bulk to boundary and bulk to bulk propagators
are very simple in momentum space. This allows us to easily perform the 
$z$-integrals that appear in transition amplitudes. 

It is convenient to work in ``axial gauge'' where we set the radial-component to $0$. In this gauge, the non-normalizable free wave-functions in AdS (these are the 
same as the bulk to boundary propagators)
 are given by
\be
A_{i} = \sqrt{2 \over \pi}  \ep_i (\norm{k} z)^{{1 \over 2}} e^{i \vect{k} \cdot \vect{x}}  K_{1 \over 2}(|\vect{k}| z) = \ep_i e^{-\norm{k} z} e^{i \vect{k} \cdot \vect{x}}; \quad  A_0 = 0; \quad \vect{\ep} \cdot \vect{k} = 0,
\ee
for $\vect{k}^2 > 0$ i.e. for spacelike momenta. For timelike momenta, the modified Bessel function $K$ should replaced by a
Hankel function of the first kind --- $H^{(1)}(\norm{k} z)$. However, it is more convenient to continue using the expressions above and simply interpret $\norm{k}$ as an imaginary quantity when $\vect{k}$ is timelike.\footnote{Strictly speaking, the identity, $H^{(1)}_{\alpha}(x) = {2 \over \pi} (-i)^{\alpha + 1} K_{\alpha}(-i x)$ tells us that we should take $\norm{k}$ to have a {\em negative} imaginary part for timelike momenta.} 

This normalization of the bulk-boundary propagators is chosen so that the 
two point function of the currents is normalized as:
\be
T( \vect{\ep_1},\vect{k_1}, \vect{\ep_2}, \vect{k_2}) =  -(2 \pi)^3 i \delta^3(\vect{k_1} - \vect{k_2}) \big(\vect{\ep_1} \cdot \vect{\ep_2}\big)  \norm{k_1}.
\ee

Below, we will also need the normalizable free wave-function of the gauge field, which is:
\be
A_{i} = \ep_i  z^{{1 \over 2}} e^{i \vect{k} \cdot \vect{x}}  J_{1 \over 2}(|\vect{k}| z); \quad  A_0 = 0; \quad \vect{\ep} \cdot \vect{k} = 0, 
\ee
and exists only for {\em timelike momenta}, $\vect{k}^2 < 0$. We caution the reader that we have normalized this solution differently from the bulk to boundary propagator.

Next, we need the Feynman rules for Yang Mills theory in AdS. For simplicity, let us consider color-ordered correlators, which correspond to color-ordered amplitudes in the bulk. 
The Feynman rules for color-ordered diagrams, as generalized to AdS, have
a three and a four point vertex, which is given by:
\begin{equation}
\label{feynmanads}
\begin{split}
V_3 &= {1 \over \sqrt{2}} \left[ \left(a_1^{\mu} \lrdel_{\nu} (a_2)_{\mu}\right) a_3^{\nu}  +  \left(a_3^{\mu} \lrdel_{\nu} (a_1)_{\mu}\right) a_2^{\nu} + 
\left(a_2^{\mu} \lrdel_{\nu} (a_3)_{\mu}\right) a_1^{\nu} \right], \\
V_4 &= i \left[(\vect{a_1 } \cdot \vect{ a_3}) (\vect{a_2 } \cdot \vect{ a_4}) - {1 \over 2} (\vect{a_1 } \cdot \vect{ a_2}) (\vect{a_3 } \cdot \vect{ a_4}) - {1 \over 2} (\vect{a_1 } \cdot \vect{ a_4}) (\vect{a_2 } \cdot \vect{ a_3})  \right],
\end{split}
\end{equation}
where the $a_n$ represent the external lines that meet at the vertex and $\vect{A} \lrdel \vect{B} \equiv \vect{A} \nabla \vect{B} - \vect{B} \nabla \vect{A}$ for two vector fields $\vect{A}$ and $\vect{B}$. The connection coefficients are given by:
\begin{equation}
\label{connectionval}
\Gamma^{\rho}_{\alpha \beta} = {1 \over 2} g^{\rho \delta} \left(\partial_{\alpha} g_{\beta \delta} + \partial_{\beta} g_{\alpha \delta} - \partial_\delta g_{\alpha \beta} \right) =  {1 \over z} \left(\delta^{\rho}_{0} \eta_{\alpha \beta} - \delta^{0}_{\alpha} \delta^{\rho}_{\beta} - \delta^{0}_{\beta} \delta^{\rho}_{\alpha} \right).
\end{equation}
Here $0$ represents the radial direction in AdS.

Below we work out the the three point transition amplitudes for the 
different combinations of helicities.
We will use the symbol $E_p$ below to mean:
\begin{equation}
\label{defep}
E_p \equiv (\norm{k_1} + \norm{k_2} + \norm{k_3}) = \norm{k_1} + \norm{k_2} + i p.
\end{equation}

First, we note that for every amplitude we have a leading factor that comes from the $z$ integrals
\begin{equation}
\label{fprefact}
\begin{split}
R^{\text{YM}}(|\vect{k_1}|, |\vect{k_2}|, p) &= { 2 \sqrt{\norm{k_1} \norm{k_2}}  \over \pi} \int_0^{\infty} z^{3 \over 2} K_{1/2}(|\vect{k_1}| z) K_{1/2}( |\vect{k_2}| z) J_{1/2} (p z) d z\\  &= \frac{\sqrt{\frac{2 p }{\pi }}}{\norm{k_1}^2+2 \norm{k_2} \norm{k_1}+\norm{k_2}^2+p^2}.
\end{split}
\end{equation}

We should remind the reader that the answers for correlators worked out in \cite{Maldacena:2011nz} involve the radial integral:
\be
\label{rympm}
\begin{split}
R^{\text{YM}}_{\text{PM}}(\norm{k_1}, \norm{k_2}, \norm{k_3}) &= \left({2 \over \pi}\right)^{3 \over 2} \sqrt{\norm{k_1} \norm{k_2} \norm{k_3}} \int_{0}^{\infty} z^{{3 \over 2}} K_{1/2}(|\vect{k_1}| z) K_{1/2}( |\vect{k_2}| z) K_{1/2} (\norm{k_3} z) d z \\ &= {1 \over \norm{k_1} + \norm{k_2} + \norm{k_3}}.
\end{split}
\ee
Apart from a normalization factor that arises because the normalizable mode in the transition amplitude is normalized differently compared to the bulk-boundary propagators, the answers for transition amplitudes and correlators are related by the following simple substitution in the term that comes
from the radial integral:
\begin{equation}
\label{substcorramp}
R^{\text{YM}}(\norm{k_1}, \norm{k_2}, \norm{p}) = {1 \over \sqrt{2 \pi p}} \left(R^{\text{YM}}_{\text{PM}}(\norm{k_1}, \norm{k_2}, -i \norm{p}) - R^{\text{YM}}_{\text{PM}}(\norm{k_1}, \norm{k_2}, i \norm{p}) \right).
\end{equation}

The tensor structures that we need to compute is given by:
\begin{equation}
T_3^* = R^{\text{YM}}(|\vect{k_1}|, |\vect{k_2}|, p) {i \over \sqrt{2}} \bigl\{ (\vect{\ep_1} \cdot \vect{\ep_2}) (\vect{k_2}- \vect{k_1})\cdot \vect{\ep_3} + (2 \vect{\ep_2} \cdot \vect{k_1}) (\vect{\ep_1} \cdot \vect{\ep_3}) - 2 (\vect{\ep_1 } \cdot \vect{ k_2}) (\vect{\ep_2} \cdot \vect{\ep_3}) \bigr\}.
\end{equation}
(Here the dependence of $T$ on the momenta and polarizations is not shown explicitly although we are using the subscript $3$ to remind the reader that this is a three-point function and the superscript $*$ to indicate that this is a transition amplitude.)
Except for the radial part, which is modified as explained above, the spinor expressions that we obtain from polarization contractions are the same as the expressions for correlators in \cite{Maldacena:2011nz}. 

\subsubsection{$++-$ Amplitude \label{subsecgaugeppm}}
We use the polarization vectors given in section \ref{secspinoreview}.  Notice that there is a leading factor of ${1 \over \sqrt{2}}$ from the interaction vertex
that gets multiplied with the norm factors from the denominator of the polarizations. We get another factor of $2$ in the denominator
when we convert momentum dot products to spinor contractions. With these observations, we see that the $++-$ amplitude is given by: 
\begin{equation}
\begin{split}
T_3^*(+,+,-) = {i R^{\text{YM}}(|\vect{k_1}|, |\vect{k_2}|, p) \over 2 \sqrt{2} |\vect{k_1}| |\vect{k_2}| p} 
\Bigl( &\dotlb[\lb_1, \lb_2]^2 \dotl[\la_2,\la_3] \dotlm[\lb_2,\la_3] + \dotlb[\lb_2, \lb_1] \dotlm[\lb_2,\la_1] \dotlm[\lb_1, \la_3]^2 \\ &-  \dotlb[\lb_1,\lb_2] \dotlm[\lb_1,\la_2] \dotlm[\lb_2,\la_3]^2 \Bigr).
\end{split}
\end{equation}

Actually each term inside the brackets is proportional to the same quantity and
we can write the whole amplitude in the flat space MHV form multiplied by a pre-factor. To see this,
we note the following relations:
\begin{equation}
\la_1 \lb_1 + \la_2 \lb_2 + \la_3 \lb_3 = i E_p \sigma^3,
\end{equation}
which leads to
\begin{equation}
\dotl[\la_1,\la_3]\dotlb[\lb_3,\lb_2] = -i E_p \dotlm[\la_1,\lb_2],
\end{equation}
and similar identities for other pairs of spinors. Moreover, we also have the identity
\begin{equation}
\begin{split}
\dotl[\la_1, \la_2] \dotlb[\lb_1,\lb_2]&= -2 \big((\vect{k_1} \cdot \vect{k_2}) -  \norm{k_1}\norm{k_2}\big) = (\norm{k_1} + \norm{k_2})^2  -  (\vect{k_1} + \vect{k_2})^2 \\ &= -i E_p (p + i( \norm{k_1} + \norm{k_2})).
\end{split}
\end{equation}
Substituting this we find that 
\begin{equation}
\label{ppm}
\begin{split}
T_3^*(+,+,-) = {-R^{\text{YM}}(|\vect{k_1}|, |\vect{k_2}|, p) \over 2 \sqrt{2}  |\vect{k_1}| |\vect{k_2}| p} &\bigl(\norm{k_2} + i p  - \norm{k_1} \bigr) \bigl(i p + \norm{k_1} - \norm{k_2} \bigr) \bigl(\norm{k_1} + \norm{k_2} - i p \bigr)
\\ &\times {\dotlb[\lb_1, \lb_2]^4 \over \dotlb[\lb_1, \lb_2] \dotlb[\lb_2, \lb_3] \dotlb[\lb_3, \lb_1]}.
\end{split}
\end{equation}

\subsubsection{$+++$ Amplitude}
The $+++$ amplitude is given by
\begin{equation}
\begin{split}
T_3^*(+,+,+) =  {i R^{\text{YM}}(|\vect{k_1}|, |\vect{k_2}|, p) \over 2 \sqrt{2} |\vect{k_1}| |\vect{k_2}| p} 
\Bigl( &\dotlb[\lb_1, \lb_2]^2 \dotlm[\la_2,\lb_3] \dotlb[\lb_2,\lb_3] + \dotlb[\lb_2, \lb_1] \dotlm[\lb_2,\la_1] \dotlb[\lb_1, \lb_3]^2 \\ &- 2 \dotlb[\lb_1,\lb_2] \dotlm[\lb_1,\la_2] \dotlb[\lb_2,\lb_3]^2 \Bigr).
\end{split}
\end{equation}
After using the identities above, we find that
\begin{equation}
\label{ppp}
T_3^*(+,+,+) =  {- R^{\text{YM}}(|\vect{k_1}|, |\vect{k_2}|, p) \over 2 \sqrt{2} |\vect{k_1}| |\vect{k_2}| p} E_p \dotlb[\lb_1, \lb_2] \dotlb[\lb_2, \lb_3] \dotlb[\lb_3, \lb_1].
\end{equation}

\subsubsection{$---$ Amplitude}
The $---$ amplitude is related to the $+++$ amplitude by parity and 
is given by
\begin{equation}
\label{mmm}
T_3^*(-,-,-) =  {-R^{\text{YM}}(|\vect{k_1}|, |\vect{k_2}|, p) \over 2 \sqrt{2}  |\vect{k_1}| |\vect{k_2}| p} E_p \dotl[\la_1, \la_2] \dotl[\la_2, \la_3] \dotlb[\la_3, \la_1].
\end{equation}

\subsubsection{$- - +$ Amplitude}
The $--+$ amplitude is related to the $++-$ amplitude by parity and 
is given by
\begin{equation}
\label{mmp}
\begin{split}
T_3^*(-,-,+) = {-R^{\text{YM}}(|\vect{k_1}|, |\vect{k_2}|, p) \over 2 \sqrt{2}  |\vect{k_1}| |\vect{k_2}| p}  &\bigl(\norm{k_2} + i  p  - \norm{k_1} \bigr) \bigl(i p + \norm{k_1} - \norm{k_2} \bigr) \bigl(\norm{k_1} + \norm{k_2} - i p \bigr) 
\\ &\times {\dotl[\la_1, \la_2]^4 \over \dotl[\la_1, \la_2] \dotl[\la_2, \la_3] \dotl[\la_3, \la_1]}.
\end{split}
\end{equation}

The list above covers all possible three point transition amplitudes. The amplitude for any other combination of helicities can be obtained by just cyclically permuting the spinor expressions, while keeping $R^{\text{YM}}$ unchanged.

\subsubsection{Flat Space Limit}
When the three point amplitudes are written in the forms above, it is manifest that the flat space limit described in \cite{Raju:2012zr} holds. Let us
remind the reader that in \cite{Raju:2012zr}, we conjectured that 
the $n$-point conserved current correlator in $3$ dimensions and the flat space gluon scattering amplitude in $4$ dimensions should be related, at tree level, through:
\be
M(\vect{\ep_1}, \vect{\tilde{k}_1}, \ldots \vect{\ep_n}, \vect{\tilde{k}_n}) = \lim_{ {\tiny{\left(\sum \norm{k_m}\right)}} \rightarrow  0}  \left(\sum \norm{k_m}\right)T(\vect{\ep_1}, \vect{k_1}, \ldots \vect{\ep_n}, \vect{k_n}).
\ee 
Here $\vect{\tilde{k}_m}$ are the on-shell $4$ dimensional vectors produced by taking the 3-dimensional vector $\vect{k_m}$ and
appending its norm to form the 4-dimensional vector $\vect{\tilde{k}_m} = \{\vect{k_m}, i \norm{k_m} \}$

Now, as we mentioned above, to compute correlators rather than transition amplitudes all we need to do is to replace $R^{\text{YM}}$ above with
 $R^{\text{YM}}_{\text{PM}}$ defined in 
\eqref{rympm}.

For example, looking at the $++-$ correlator (the $--+$ case works in 
exactly the same way), we see that:
\be
T_3(+,+,-) \underset{E_p \rightarrow 0}{\longrightarrow}  i { 2 \sqrt{2} \over E_p} {\dotlb[\lb_1, \lb_2]^4 \over \dotlb[\lb_1, \lb_2] \dotlb[\lb_2, \lb_3] \dotlb[\lb_3, \lb_1]}.
\ee
This is because at $E_p = 0$, we can replace ${\norm{k_1} + \norm{k_2} - i p \over p} = -2 i$ and so the numerator in $R^{\text{YM}}_{\text{PM}}$ neatly 
cancels with the denominator leaving behind the factor of ${1 \over E_p}$. Of course this is multiplied with the famous 3-pt gluon amplitude in four dimensions, which is precisely what we expect from our flat-space conjecture.\footnote{The careful reader might note that we have an extra factor of $2 \sqrt{2}$. This is present because our polarization vectors are unconventionally normalized as shown in \eqref{normalizationpol} so that $\vect{\ep^{+}}(\vect{k}) \cdot \vect{\ep^{-}}(\vect{k}) = 2$. This normalization is convenient because below we will extend momentum vectors by their polarizations and this helps remove factors of $\sqrt{2}$ there; however these factors sometimes reappear in final results as above.}
 
\subsection{Gravity}
We now turn to the computation of three point transition amplitudes in the pure
gravity theory, using the Hilbert action in the bulk.  Just as we found above, we find that the answers for transition amplitudes are very similar to the answers for correlators, except that the part of the answer that comes from the 
radial integral over the bulk-boundary propagators gets modified as in \eqref{substcorramp}.  

The bulk to boundary propagator for gravity is given by the 
expression:
\be
\label{solgravfree}
h_{i j}(\vect{e}, \vect{k}, \vect{x},z) =  {e_{i j} \over z^2} (\norm{k} z)^{{d \over 2}} e^{i \vect{k} \cdot \vect{x}} {\sqrt{2 \over \pi}} K_{3 \over 2}(|\vect{k}| z)  = {e_{i j} \over z^2} (1 + \norm{k} z)e^{-\norm{k} z} e^{i \vect{k} \cdot \vect{x}}
\ee
It is important to note that in \eqref{solgravfree}, both indices on
$h$ are lowered. If one index had been raised the leading factor of ${1 \over z^2}$ would be absent. As explained above, this form of the bulk
to boundary propagator is correct for spacelike momentum $\vect{k} \cdot \vect{k} > 0$. For timelike momentum, we should analytically continue the
expression above while taking $\norm{k}$ to have a negative imaginary part. 

When we refer to the ``normalizable mode'' that enters transition amplitudes, we are referring to the solution:
\be
h_{i j}(\vect{e}, \vect{k}, \vect{x}, z) =  {1 \over z^2} e_{i j} z^{{3 \over 2}} J_{3 \over 2}(|\vect{k}| z) e^{i \vect{k} \cdot \vect{x}}. 
\ee
This is because it is this term that naturally enters the bulk to bulk propagator. We will have to be careful about these different normalizations when we compare correlators to transition amplitudes below. 

\subsubsection{Interaction Vertices}
To obtain the three point gravity transition amplitude, we first need to expand
the Hilbert action out to third order in fluctuations.\footnote{It is possible to 
obtain the three-point on shell amplitude without going through this process, and by using the flat
space result, as was done in \cite{Maldacena:2002vr}. Our approach is more direct. It also has the advantage that it helps us keep track of the boundary terms that we are adding to the action. It also lets us see how the on-shell computation is much simpler than using Feynman rules.} 
The computations in this subsection were performed using the excellent program Xact \cite{xactwebsite,martin2008invar,martin2007invar,brizuela2009xpert} that allowed us to automate the tensor manipulations below. 

We start with the cubic action given in \cite{Arutyunov:1999nw}
\be
\label{cubicgravaction1}
S = {-1 \over 6} \int \sqrt{|g|}(V_{\mu \nu} - {1 \over 2} g_{\mu \nu} V) h^{\mu\nu},
\ee
where 
\be
\begin{split}
V^{\mu \nu} &= 
 -\nabla_{\rho}\left(h^{\rho \sigma} (\nabla^{\mu} h^{\nu}_{\sigma}  + 
      \nabla^{\nu} h^{\mu}_{\sigma}  -  
      \nabla_{\sigma} h^{\mu \nu}) \right)  + 
 \nabla^{\nu}  \left( h^{\rho \sigma} \nabla^{\mu} h_{\rho \sigma} \right) \\ &+  {1 \over 2}
     (\nabla^{\mu} h^{\nu}_{\rho}  + \nabla^{\nu} h^{\mu}_{\rho}  - 
    \nabla_{\rho} h^{\mu \nu} ) \nabla^{\rho}
   h^{a b} g_{a b}  - {1 \over 2} \nabla^{\mu}
   h_{\rho \sigma}  \nabla^{\nu} h^{\rho \sigma}  + 
 \nabla_{\rho} h^{\mu}_{\sigma}  \nabla^{\rho} h^{\sigma \nu}  \\ &- 
 \nabla_{\sigma} h^{\mu}_{\rho}  \nabla^{\rho} h^{\sigma \nu}.
\end{split}
\ee

This term still contains various terms with a double derivative. To remove
these we need to add the two-derivative terms
\be
\begin{split}
B = &-\frac{1}{2} g^{cd} g_{\mu \nu }
   \nabla_{d}\left(h^{\mu \nu } h^{\rho \sigma }
   \nabla_{c}{h_{\rho \sigma
   }}\right)+\nabla_{\nu }\left(h^{\mu \nu } h^{\rho
   \sigma } \nabla_{\mu }{h_{\rho \sigma
   }}\right)+\frac{1}{2} g^{cd} g_{\mu \nu }
   \nabla_{\rho }\left(h^{\mu \nu } h^{\rho \sigma }
   \nabla_{c}{h_{d\sigma }}\right) \\ 
&+\frac{1}{2}
   g^{cd} g_{\mu \nu } \nabla_{\rho }\left(h^{\mu
   \nu } h^{\rho \sigma }
   \nabla_{d}{h_{c\sigma }}\right)-\nabla_{\rho
   }\left(h^{\mu \nu } h^{\rho \sigma } \nabla_{\mu
   }{h_{\nu \sigma }}\right)-  \nabla_{\rho } \left(h^{\mu \nu } h^{\rho
   \sigma } \nabla_{\nu}{h_{\mu \sigma }}\right) \\ &-\frac{1}{2} g^{cd} g_{\mu
   \nu } \nabla_{\rho }\left(h^{\mu \nu } h^{\rho
   \sigma } \nabla_{\sigma
   }{h_{cd}}\right)+\nabla_{\rho }\left(h^{\mu \nu }
   h^{\rho \sigma } \nabla_{\sigma }{h_{\mu
   \nu }}\right).
\end{split}
\ee
Here, we should note that although in position space these terms can only 
give a delta function contribution to boundary correlators, we could have been worried about them in
momentum space. This is because the recursion relations involve multiplying
two three-point functions in momentum space, or convoluting two three-point
functions in position space; in this manner what was a delta function contribution may become important. However, fortuitously, when we evaluate this
boundary term on linearized solutions to the equations of motion 
in the gauge
\be
\label{gaugechoice}
h^{0 \mu} = 0, \quad h^{\mu}_{\mu} = 0,
\ee
and {\em on shell}, so that
\be 
\label{onshell}
\partial_{\mu} h^{\mu \nu} = 0,
\ee
the boundary terms in $B$ genuinely vanish upon integration.
We find then that
\be
\begin{split}
(V_{\mu \nu} &- {1 \over 2} g_{\mu \nu} V) h^{\mu\nu} - B = 
\frac{1}{2} g^{cd} g_{\mu \nu } h^{\rho \sigma } \nabla_{c}h_{\rho \sigma } \nabla_{d}h^{\mu \nu } + \frac{1}{4} g^{cd} g_{\mu \nu } h^{\mu \nu } \nabla_{c}h_{\rho \sigma } \nabla_{d}h^{\rho \sigma } \\ &-  h^{\rho \sigma } \nabla_{\mu }h_{\rho \sigma } \nabla_{\nu }h^{\mu \nu } - \frac{1}{2} h^{\mu \nu } \nabla_{\mu }h_{\rho \sigma } \nabla_{\nu }h^{\rho \sigma } - \frac{1}{2} g^{cd} g_{\mu \nu } h^{\rho \sigma } \nabla_{c}h_{d\sigma } \nabla_{\rho }h^{\mu \nu } \\ &- \frac{1}{2} g^{cd} g_{\mu \nu } h^{\rho \sigma } \nabla_{d}h_{c\sigma } \nabla_{\rho }h^{\mu \nu } + h^{\rho \sigma } \nabla_{\mu }h_{\nu \sigma } \nabla_{\rho }h^{\mu \nu } -  h^{\mu \nu } \nabla_{\nu }h_{\mu \sigma } \nabla_{\rho }h^{\rho \sigma } \\ &- \frac{1}{4} g_{ab} g^{cd} g_{\mu \nu } h^{\mu \nu } \nabla_{c}h_{d\rho } \nabla^{\rho }h^{ab} - \frac{1}{4} g_{ab} g^{cd} g_{\mu \nu } h^{\mu \nu } \nabla_{d}h_{c\rho } \nabla^{\rho }h^{ab} \\ &+ \frac{1}{2} g_{ab} h^{\mu \nu } \nabla_{\mu }h_{\nu \rho } \nabla^{\rho }h^{ab} + \frac{1}{2} g_{ab} h^{\mu \nu } \nabla_{\nu }h_{\mu \rho } \nabla^{\rho }h^{ab} + \frac{1}{4} g_{ab} g^{cd} g_{\mu \nu } h^{\mu \nu } \nabla_{\rho }h_{cd} \nabla^{\rho }h^{ab} \\ &- \frac{1}{2} g_{ab} h^{\mu \nu } \nabla_{\rho }h_{\mu \nu } \nabla^{\rho }h^{ab} - \frac{1}{2} g^{cd} g_{\mu \nu } h^{\mu \nu } \nabla_{\rho }h_{c\sigma } \nabla^{\rho }h^{\sigma }{}_{d} + h^{\mu \nu } \nabla_{\rho }h_{\mu \sigma } \nabla^{\rho }h^{\sigma }{}_{\nu } \\ &+ \frac{1}{2} g^{cd} g_{\mu \nu } h^{\rho \sigma } \nabla_{\rho }h^{\mu \nu } \nabla_{\sigma }h_{cd} + \frac{1}{2} g^{cd} g_{\mu \nu } h^{\mu \nu } \nabla^{\rho }h^{\sigma }{}_{d} \nabla_{\sigma }h_{c\rho } -  h^{\rho \sigma } \nabla_{\rho }h^{\mu \nu } \nabla_{\sigma }h_{\mu \nu }\\ &-  h^{\mu \nu } \nabla^{\rho }h^{\sigma }{}_{\nu } \nabla_{\sigma }h_{\mu \rho }.
\end{split}
\ee

We write this term out in gory detail because this is what we would have to 
use if we were to try and compute an exchange Feynman diagram. Fortunately,
in our method we only need the {\em on-shell} three point function. 
When we impose just the traceless conditions from \eqref{gaugechoice},
we 
find a remarkable simplification:
\be
\begin{split}
(V_{\mu \nu} - {1 \over 2} g_{\mu \nu} V) h^{\mu\nu} - B = - h^{\mu \nu } \bigl(&{3 \over 2} \nabla_{\mu }h^{\rho \sigma } \nabla_{\nu }h_{\rho \sigma } + 
+ \nabla^{\rho }h_{\mu \nu } \nabla_{\sigma }h_{\rho }{}^{\sigma } 
\\ &-  2 \nabla_{\nu }h_{\rho \sigma } \nabla^{\sigma }h_{\mu }{}^{\rho } + \nabla_{\rho }h_{\nu \sigma } \nabla^{\sigma }h_{\mu }{}^{\rho } -  \nabla_{\sigma }h_{\nu \rho } \nabla^{\sigma }h_{\mu }{}^{\rho }\bigr).
\end{split}
\ee

We now convert all covariant derivatives to partial derivatives using the 
connection coefficients \eqref{connectionval}, and again impose
the tracelessness condition on $h$. This leads to 
\begin{align}
\label{linewithdivergence}
(V_{\mu \nu} &- {1 \over 2} g_{\mu \nu} V) h^{\mu\nu} - B = 4 h_{a}{}^{c} h^{ab} h_{bc} - \frac{3}{2} h^{ab} \partial_{a}h^{cd} \partial_{b}h_{cd} -h^{a b} \partial^{c} h_{a b} \partial_d h_c^d \\ &+ 2 h^{ab} \partial_{b}h_{cd} \partial^{d}h_{a}{}^{c}  
- h^{ab} \partial_{c}h_{bd} 
\partial^{d}h_{a}{}^{c} 
 + h^{ab} \partial_{d}h_{bc} \partial^{d}h_{a}{}^{c} + 2 z 
h_{a}{}^{c} h^{ab} \partial_{0}h_{bc}.
\end{align}
If we now also use the on-shell condition $\partial_{a} h^{a b} = 0$, then
the third term in the first line above (Eqn. \eqref{linewithdivergence})
drops out. This results in
\be
\begin{split}
(V_{\mu \nu} - {1 \over 2} g_{\mu \nu} V) h^{\mu\nu} - B = &4 h_{a}{}^{c} h^{ab} h_{bc} - \frac{3}{2} h^{ab} \partial_{a}h^{cd} \
\partial_{b}h_{cd} + 2 h^{ab} \partial_{b}h_{cd} 
\partial^{d}h_{a}{}^{c} \\ &-  h^{ab} \partial_{c}h_{bd} 
\partial^{d}h_{a}{}^{c}   
 + h^{ab} \partial_{d}h_{bc} \partial^{d}h_{a}{}^{c} + 2 z 
h_{a}{}^{c} h^{ab} \partial_{0}h_{bc}.
\end{split}
\ee
By adding another total derivative term, which vanishes on-shell we find
\be
\begin{split}
(V_{\mu \nu} - {1 \over 2} g_{\mu \nu} V) h^{\mu\nu} &- B + \partial_c \Big(h^{a b} h_{b d} \partial_d h^c_a \Big) = 4 h_{a}{}^{c} h^{ab} h_{bc} - \frac{3}{2} h^{ab} \partial_{a}h^{cd} \partial_{b}h_{cd} \\ &+ 3 h^{ab} \partial_{b}h_{cd} \partial^{d}h_{a}{}^{c}   + h^{ab} \partial_{d}h_{bc} \partial^{d}h_{a}{}^{c} + 2 z h_{a}{}^{c} h^{ab} \partial_{0}h_{bc}.
\end{split}
\ee

We can make this even simpler and get rid of the $z$ derivatives, if we remember that in \eqref{cubicgravaction1}, 
this term is multiplied by $\sqrt{-g} = {1 \over z^{d+1}}$.  Now, 
\be
\begin{split}
&{1 \over z^{d +1}} \left(4 h_{a}{}^{c} h^{ab} h_{bc} +  h^{ab} \partial_{d}h_{bc} \partial^{d}h_{a}{}^{c} + 2 z h_{a}{}^{c} h^{ab} \partial_{0}h_{bc} \right)\\
&={1 \over z^{d +1}} \left(4 h_{a}^{c} h^{a}_{b} h^{b}_{c} +  z^2 \eta^{i j} h^{a}_{b} \partial_{i} \left(h^{b}_{c}  \right) \partial_{j} h_{a}^{c} +  z^4  h_{a}^{b} \partial_{0} \left({h_{b}^{c} \over z^2} \right) \partial_{0} {h^{a}_{c}} +  
2 z^3 h_{a}^{c} h^{a}_{b} \partial_{0}\left({h^{b}_{c} \over z^2} \right) \right) \\
&= {1 \over z^{d +1}} \left( z^2 \eta^{i j} h^{a}_{b} \partial_{i} \left(h^{b}_{c}  \right) \partial_{j} h_{a}^{c} +  {1 \over 2} z^2 \partial_{0}  \left(h_{a}^{b}  h_{b}^{c} \partial_{0} {h^{a}_{c}} \right) - {1 \over 2} z^2 h_{a}^{b} h_{b}^{c} \partial_0^2 h^{a}_{c}  \right) \\
&= {1 \over 2 z^{d - 1}} \eta^{i j} \partial_i \left(h^{a}_{b}  h^{b}_{c}   \partial_{j} h_{a}^{c}\right)  +  {1 \over 2 z^{d - 1}}h^{a}_{b}  h^{b}_{c}  \eta^{i j} \partial_i \partial_{j} h_{a}^{c} +  {1 \over 2} \partial_0  \left({1 \over z^{d - 1}} h_{a}^{b}  h_{b}^{c} \partial_{0} {h^{a}_{c}} \right)  \\ &\hphantom{=} - {1 \over 2} h_{a}^{b} h_{b}^{c} \partial_0 \left( {1 \over z^{d - 1}} \partial_0 h^{a}_c \right).
\end{split}
\ee
However, on shell, $h$ precisely satisfies the equation (see, for example, the detailed review of perturbation theory
in \cite{Raju:2011ed})
\be
 {1 \over z^{d - 1}}h^{a}_{b}  h^{b}_{c}  \eta^{i j} \partial_i \partial_{j} h_{a}^{c}   - {1 \over 2} h_{a}^{b} h_{b}^{c} \partial_0 \left( {1 \over z^{d - 1}} \partial_0 h^{a}_c \right) = 0.
\ee
After recalling the factor of ${-1 \over 6}$ in \eqref{cubicgravaction1},this
leads to the following expression for the three point function.
\be
\begin{split}
\label{gravity3ptvertex}
&T(\vect{e_1}, \vect{k_1},\vect{e_2}, \vect{k_2}, \vect{e_3}, \vect{k_3}) = \sum_{\pi} \int {F_{\pi} \over z^{d-1}} \\  
&\begin{split}
&F_{\pi}  =\Big({1 \over 4} \dotp[\ep_{\pi_{1}}, k_{\pi_{2}}] \dotp[\ep_{\pi_{1}}, k_{\pi_{3}}] \dotp[\ep_{\pi_{2}}, \ep_{\pi_{3}}]^2 - {1 \over 2} \dotp[\ep_{\pi_{1}}, k_{\pi_{2}}] \dotp[\ep_{\pi_{1}}, \ep_{\pi_{3}}] \dotp[\ep_{\pi_{2}}, k_{\pi_{3}}] \dotp[\ep_{\pi_{2}}, \ep_{\pi_{3}}]\Big) \\ 
& \hphantom{F_{\pi}  =\Big(}
 \times \phi(|\vect{k_1}| z)  \phi(|\vect{k_2}| z)  \phi(|\vect{k_3}| z), \end{split}
\end{split}
\ee
where $\pi$ runs over the permutation group of 3 elements and $\phi$ is the
radial part of the wave-function defined by
\be
h^i_j(\vect{e}, \vect{k}, \vect{x},z) \equiv z^2 \eta^{i l} \ep_{l} \ep_{j} e^{i \vect{k} \cdot \vect{x}} \phi(|\vect{k}| z),
\ee
with the linearized solutions $h^{i}_{j}$ defined in \eqref{solgravfree}. Note
that $\phi$ carries information about whether any of the wave functions
we are using is normalizable.

\subsubsection{Answers for Three Point Stress Tensor Transition Amplitudes}
Let us now compute the three-point transition amplitudes of the stress-tensor that we need to compute 4-pt correlators. First we need the radial integral; 
this gives a polarization-independent part of the answer, 
which is then multiplied by some function that depends on the polarizations.

The radial integral gives us
\be
\begin{split}
R^{\text{gr}}(|\vect{k_1}|, |\vect{k_2}|, p) &= \int  \phi(|\vect{k_1}| z)  \phi(|\vect{k_2}| z)  \phi(|\vect{k_3}| z) {d z \over z^2} \\ &= {2 \over \pi} \big(\norm{k_1} \norm{k_2} \big)^{3 \over 2} \int {d z \over z^2} z^{3 \over 2} K_{3 \over 2}(|\vect{k_1}| z) z^{3 \over 2} K_{3 \over 2}(|\vect{k_2}| z) z^{3 \over 2} J_{3 \over 2}(p z)\\ &= \frac{p^{3/2} \left(\norm{k_1}^2+4 \norm{k_2} \norm{k_1}+\norm{k_2}^2+p^2\right) \sqrt{\frac{2}{\pi }}}{\left(\norm{k_1}^2+2 \norm{k_2} \norm{k_1}+\norm{k_2}^2+p^2\right)^2}.
\end{split}
\ee
The pre-factor that enters the correlator was calculated in \cite{Maldacena:2011nz} and is given by:
\be
\begin{split}
\label{rgrpm}
&R_{\text{PM}}^{\text{gr}}(|\vect{k_1}|, |\vect{k_2}|, |\vect{k_3}|) =  \big({2 \over \pi} \norm{k_1} \norm{k_2} \norm{k_3} \big)^{3 \over 2}\int_{0}^{\infty} {1 \over z^2} z^{3 \over 2} K_{3 \over 2}(|\vect{k_1}| z) z^{3 \over 2} K_{3 \over 2}(|\vect{k_2}| z) z^{3 \over 2} K_{3 \over 2}(\norm{k_3} z) \\
&=  -\frac{|\vect{k_{2}}| |\vect{k_{3}}|
   |\vect{k_{1}}|}{(|\vect{k_{1}}|+|\vect{k_{2}}|+|\vect{k_{3}}|)^2}+|\vect{k_{1}}|+|\vect{k_{2}}|+
   |\vect{k_{3}}|-\frac{|\vect{k_{1}}| |\vect{k_{2}}|+|\vect{k_{3}}| |\vect{k_{2}}|+|\vect{k_{1}}|
   |\vect{k_{3}}|}{|\vect{k_{1}}|+|\vect{k_{2}}|+|\vect{k_{3}}|}.
\end{split}
\ee
The radial integral for the transition amplitude is convergent but, in the case of the correlator, it is divergent; the value in \eqref{rgrpm} comes from cutting it off at $z = \epsilon$ and picking up the $\epsilon^0$ piece. In this respect, transition amplitudes
are nicer than correlators. We discuss this in some more detail below. 
 
Apart from an overall normalization, which appears because the normalizable mode is normalized differently from the bulk-boundary propagator, note that 
the answer for the radial integral that enters the transition amplitude is closely related to the term that enters the correlator:
\be
R^{\text{gr}}(|\vect{k_1}|, |\vect{k_2}|, p) = -i \sqrt{2 \over \pi} p^{-3 \over 2}     \Big(R_{\text{PM}}^{\text{gr}}(\vect{k_1}, \vect{k_2}, i p) - R_{\text{PM}}^{\text{gr}}(|\vect{k_1}|, |\vect{k_2}|, -i p) \Big).
\ee

We adopt the same notation as \eqref{defep} for the variable $E_p$. With this definition, we find that, except for the part the comes
from the radial integrals,  the graviton transition amplitudes can be written
as the ``square'' of the gauge boson amplitudes explored in the previous subsection. 

\paragraph{$++-$ Amplitude \\}
Let us consider the first term in the sum over permutations of 
\eqref{gravity3ptvertex}, and take particles $1$ and $2$ to have positive
helicity, and particle $3$ to have negative helicity. Then we find
\be
\begin{split}
F_1 &={ -R^{\text{gr}}(k_1, k_2, p) \over  16 \norm{k_1}^2 \norm{k_2}^2 p^2}  \dotl[\lb_1, \lb_2] 
\dotlm[\lb_1, \la_2] \dotlm[\lb_1, \la_3] \dotlm[\lb_2, \la_3]^3 \\ &\times \left({-1 \over 4}  \dotl[\lb_1, \lb_3] \dotlm[\lb_2, \la_3] - {1 \over 2} \dotlm[\lb_1, \la_3] \dotl[\lb_2, \lb_3] \right) .
\end{split}
\ee
Then, after summing over permutations and using the identities of \eqref{subsecgaugeppm} we find that
the $++-$ amplitude is given by:
\begin{equation}
\label{ppmstress}
\begin{split}
T_3^*(+,+,-) = {R^{\text{gr}}(|\vect{k_1}|, |\vect{k_2}|, p) \over 32  |\vect{k_1}|^2 |\vect{k_2}|^2 p^2} \bigl(\norm{k_2} + i p  - \norm{k_1} \bigr)^2 \bigl(i p + \norm{k_1} - \norm{k_2} \bigr)^2 \bigl(\norm{k_1} + \norm{k_2} - i p \bigr)^2&\\
\times \left({\dotlb[\lb_1, \lb_2]^4 \over \dotlb[\lb_1, \lb_2] \dotlb[\lb_2, \lb_3] \dotlb[\lb_3, \lb_1]}\right)^2&
\end{split}
\end{equation}

\paragraph{$+++$ Amplitude \\}
The $+++$ amplitude is given by
\begin{equation}
\label{pppstress}
T_3^*(+,+,+) =  {R^{\text{gr}}(|\vect{k_1}|, |\vect{k_2}|, p) \over 32 |\vect{k_1}|^2 |\vect{k_2}|^2 p^2} E_p^2 \left(\dotlb[\lb_1, \lb_2] \dotlb[\lb_2, \lb_3] \dotlb[\lb_3, \lb_1]\right)^2.
\end{equation}

\paragraph{$---$ Amplitude \\}
The $---$ amplitude is related to the $+++$ amplitude by parity and 
is given by
\begin{equation}
\label{mmmstress}
T_3^*(-,-,-) =  {R^{\text{gr}}(|\vect{k_1}|, |\vect{k_2}|, p) \over 32 |\vect{k_1}|^2 |\vect{k_2}|^2 p^2} E_p^2 \left(\dotl[\la_1, \la_2] \dotl[\la_2, \la_3] \dotl[\la_3, \la_1]\right)^2.
\end{equation}

\paragraph{$- - +$ Amplitude \\}
The $--+$ amplitude is related to the $++-$ amplitude by parity and 
is given by
\begin{equation}
\label{mmpstress}
\begin{split}
T_3^*(-,-,+) = {R^{\text{gr}}(|\vect{k_1}|, |\vect{k_2}|, p) \over 32  |\vect{k_1}|^2 |\vect{k_2}|^2 p^2} \bigl(\norm{k_2} + i p  - \norm{k_1} \bigr)^2 \bigl(i p + \norm{k_1} - \norm{k_2} \bigr)^2 \bigl(\norm{k_1} + \norm{k_2} - i p \bigr)^2 &\\
\times \left({\dotl[\la_1, \la_2]^4 \over \dotl[\la_1, \la_2] \dotl[\la_2, \la_3] \dotl[\la_3, \la_1]}\right)^2&
\end{split}
\end{equation}

One point that might cause some confusion is the asymmetry between $p$ and $-p$ 
in the formulae above. This asymmetry arises from the choice of polarization-vectors, which require us to choose a sign for the norm of the third momentum. However, recall that in obtaining
the four-point function we always sum over the polarizations of the intermediate
state, and in doing this, the asymmetry will disappear.

\paragraph{No divergences from the boundary\\}
A common property of AdS/CFT correlators in momentum space is that they must
be regulated to get rid of divergences from the boundary at $z = 0$. This is true of the correlation function computations of \cite{Maldacena:2011nz} and also of \eqref{rgrpm}. However, it is remarkable that simple power counting tells us that the radial integrals
that enter transition amplitudes are convergent. 

This is clear from \eqref{gravity3ptvertex}. A non-normalizable wavefunction
behaves like $z^0$ near the boundary (this is true provided one index is raised and another is lowered), while the normalizable mode behaves like $z^3$. Although we would have obtained a ${1 \over z^4}$ from the $\sqrt{-g}$ factor, we also get one factors of $z^2$ that comes from the inverse metric
 required to contract the derivatives in the interaction vertex. Consequently,
the integrand $F_{\pi}$ in \eqref{gravity3ptvertex} goes like $z$ near the
boundary, and leads to a convergent integral. 

This removes a possible complication in using the recursion relations. The
computation of the four-point function involves the product of two transition amplitudes. So, naively one might have worried a ${1 \over \ep}$ term from one amplitude could have combined with a $\ep$ term from another amplitude to give a finite contribution. 
there are no ${1 \over \ep}$ terms at all.

\subsubsection{Flat Space Limit}
In \cite{Raju:2012zr}, we conjectured that the correlation function of the 
stress tensor and the graviton amplitude should be related, at tree level,  through
\be
M(\vect{e_1}, \vect{\tilde{k}_1}, \ldots \vect{e_n}, \vect{\tilde{k}_n}) = \lim_{\tiny{\sum \norm{k_m}} \rightarrow  0} {\left(\sum \norm{k_m}\right)^{n-1}  \over \left(\prod \norm{k_m} \right) \Gamma(n-1) } T(\vect{e_1}, \vect{k_1}, \ldots \vect{e_n}, \vect{k_n}).
\ee
Just as in the conserved-current case $\vect{\tilde{k}_m}$ are on-shell $4$ dimensional vectors produced related to the
3-dimensions vectors through $\vect{\tilde{k}_m} = \{\vect{k_m}, i \norm{k_m} \}$

To compute correlators rather than transition amplitudes all we need to do is to replace $R^{\text{gr}}$ by $R^{\text{gr}}_{\text{PM}}$ defined in \eqref{rgrpm}.
With this replacement, let us consider the $++-$ stress tensor correlator. We see that when we take $E_p \rightarrow 0$, the correlator becomes:
\be
T^{++-} = 2 {\norm{k_1} \norm{k_2} \norm{k_3}  \over  E_p^2} \left({\dotlb[\lb_1, \lb_2]^4 \over \dotlb[\lb_1, \lb_2] \dotlb[\lb_2, \lb_3] \dotlb[\lb_3, \lb_1]}\right)^2,
\ee
which is consistent with our conjecture.

On the other hand, if we consider the $T^{+++}$ correlator, then we find that it has no pole at $E_p = 0$ at all, which reflects that fact that the all-plus
graviton scattering amplitude vanishes for the Hilbert action.

Note that as we explained in \cite{Raju:2012zr}, flat space S-matrix
elements can be extracted even if we go beyond the Hilbert action. For example
if we compute correlation functions using a $W^3$ action in flat space, this
gives a non-zero all-plus scattering amplitude. Corresponding to this, the $W^3$ action gives a correlator in AdS$_4$ that has a singularity of order $E_p^6$. (See Equation 2.18 of \cite{Maldacena:2011nz}.)

\section{Formulas for Four Point Functions \label{secformula}}
We will now use the recursion relations developed in \cite{Raju:2012zr}
to write down formulas for the four point functions of stress tensors
and currents. This process proceeds in the following steps. First, we describe, in detail, a one-parameter deformation of each external momentum by a null vector with the property that it preserves the norm of them momentum.   The analytic properties of the correlator under this extension can be used
to obtain recursion relations for the four point function in terms of 
deformed three point functions; we describe this procedure next. 
In the next section we evaluate these formulas for correlators of both
the stress tensor and conserved currents and  verify that the answers
 have the correct flat
space limit. 

\subsection{Extending the Momenta}
For the four point function, we start by deforming all four momenta through
\begin{equation}
\label{risagerextension}
\vect{k_m} \rightarrow \vect{k_m} + \alpha_m \vect{\ep_m} w, 
\end{equation}
where there is {\em no sum} on the $m$ in the second term. The four $\alpha$'s are fixed by the equation
\be
\label{constraint}
\sum_{p=1}^4 \alpha_{p} \vect{\ep_{p}} = 0.
\ee
This has a unique solution up to one complex multiplicative parameter that
can be absorbed in the definition of $w$. 

In fact the extension \eqref{risagerextension} can be conveniently
rephrased in terms of spinors. For each momentum, only one of the spinors --- either $\la_m$ or $\lb_m$ is extended --- as shown below, where we use the notation $\beta_m = {\alpha_m \over i \norm{k_m}}$. 
\be
\begin{array}{lll}
\text{negative polarization:}& \la_m(w) = \la_m; & \lb_m(w) = \lb_m +  \beta_m \lad_m w; \\
\text{positive polarization:} & \la_m(w) = \la_m + \beta_m \lbd_m w; 
 \quad \quad & \lb_m(w) = \lb_m.  \\
\end{array}
\ee

\paragraph{Explicit Expressions for $\beta_m$}
It is quite easy to find explicit expressions for the $\beta_m$ given 
a set of external helicities. We enumerate these expressions for 
different possible external helicities.
\begin{enumerate}
\item{ ${\bf \{h_1, h_2, h_3, h_4\} = \{1,-1,1,-1\}}$}\\
In terms of the $\beta_m$ variables we have the equations:
\begin{equation}
\beta_1 \lbd_1 \lb_1 + \beta_2 \la_2 \lad_2 + \beta_3 \lbd_3 \lb_3 + \beta_4 \la_4 \lad_4 = 0.
\end{equation}
Dotting this equation with $\la_4 \lb_3$, and then $\la_2 \lb_3$ and then $\la_2 \lad_4$ we find that
\begin{equation}
\label{explicitbeta}
\begin{split}
{\beta_2 \over \beta_1} &= {- \dotlm[\lb_1, \la_4] \dotlb[\lb_1, \lb_3] \over \dotl[\la_2, \la_4] \dotlm[\la_2, \lb_3]}; ~
{\beta_3 \over \beta_1}  = {-\dotlm[\la_2, \lb_1] \dotlm[\la_4,\lb_1] \over \dotlm[\la_2,\lb_3] \dotlm[\la_4, \lb_3]}; ~
{\beta_4 \over \beta_1} = {-\dotlm[\lb_1, \la_2] \dotl[\lb_1, \lb_3] \over \dotl[\la_4, \la_2] \dotlm[\la_4, \lb_3]}.
\end{split}
\end{equation}
It is this combination of helicities that we will use in sections \ref{secmhvym} and \ref{secmhvgrav}, and all appearances of $\beta_m$ in that section refer to the quantities defined above.
\item{ ${\bf \{h_1, h_2, h_3, h_4\} = \{1,1,1,-1\}}$}\\
We now have the equations
\begin{equation}
\beta_1 \lbd_1 \lb_1 + \beta_2 \lbd_2 \lb_2 + \beta_3 \lbd_3 \lb_3 + \beta_4 \la_4 \lad_4 = 0.
\end{equation}
This leads to 
\be
{\beta_2 \over \beta_1} = -{\dotl[\lb_1, \lb_3] \dotlm[\la_4, \lb_1] \over \dotl[\lb_2, \lb_3] \dotlm[\la_4, \lb_2]};~ 
{\beta_3 \over \beta_1} = -{\dotl[\lb_1, \lb_2] \dotlm[\la_4, \lb_1] \over \dotl[\lb_3, \lb_2] \dotlm[\la_4, \lb_3]}; ~
{\beta_4 \over \beta_1} = -{\dotl[\lb_1, \lb_2] \dotl[\lb_1, \lb_3] \over \dotlm[\la_4, \lb_2] \dotlm[\la_4, \lb_3]}.
\ee
\item {${\bf \{h_1, h_2, h_3, h_4\} = \{1,1,1,1\}}$}
This gives rise to 
\begin{equation}
\beta_1 \lbd_1 \lb_1 + \beta_2 \lbd_2 \lb_2 + \beta_3 \lbd_3 \lb_3 + \beta_4 \lbd_4 \lb_4 = 0,
\end{equation}
which has the solution
\be
{\beta_2 \over \beta_1} = -{\dotl[\lb_1, \lb_4] \dotl[\lb_1, \lb_3] \over \dotl[\lb_2, \lb_4] \dotl[\lb_2, \lb_3]};~
{\beta_3 \over \beta_1} = -{\dotl[\lb_1, \lb_4] \dotl[\lb_1, \lb_2] \over \dotl[\lb_3, \lb_4] \dotl[\lb_3, \lb_2]}; ~ 
{\beta_4 \over \beta_1} = -{\dotl[\lb_1, \lb_2] \dotl[\lb_1, \lb_3] \over \dotl[\lb_4, \lb_2] \dotl[\lb_4, \lb_3]}. 
\ee
\end{enumerate}
All other helicity combinations can be obtained through interchanges in the expressions above or using parity. 

\subsection{Recursion Relations}
With this extension, we can now write down 
the recursion relations derived in \cite{Raju:2012zr}. However, when the boundary dimension is odd, as it is in this case, we find an important simplification. To see this we rewrite the expression for the bulk to bulk propagator 
given in \cite{Raju:2010by,Raju:2011ed} and used in  \cite{Raju:2012zr} as follows:
\begin{equation}
\label{axialpropagatorinf}
\begin{split}
&G^{\text{axial},{\rm a  b}}_{i j}(\vect{k}, z, z') = \int_0^{\infty} {-i  p d p }   \Bigl[{(z z')^{1 \over 2} J_{1 \over 2}(p z) J_{1 \over 2} (p z') {\cal T}_{i j} \delta^{a b}\over 
\left(\vect{k}^2 + p^2 - i \epsilon \right)}\Bigr] \\ 
&= \int_{0}^{\infty} {-2 i  d p \over \pi }  \Bigl[{\sin(p z) \sin (p z') {\cal T}_{i j} \delta^{a b}\over 
\left(\vect{k}^2 + p^2 - i \epsilon \right)}\Bigr] 
= \int_{-\infty}^{\infty} {-i  d p \over \pi }  \Bigl[{\sin(p z) \sin (p z') {\cal T}_{i j} \delta^{a b}\over 
\left(\vect{k}^2 + p^2 - i \epsilon \right)}\Bigr] \\
&= {1 \over 2} \int_{-\infty}^{\infty} {-i  p d p }   \Bigl[{(z z')^{1 \over 2} J_{1 \over 2}(p z) J_{1 \over 2} (p z') {\cal T}_{i j} \delta^{a b}\over 
\left(\vect{k}^2 + p^2 - i \epsilon \right)}\Bigr]. \\
\end{split}
\end{equation}
Here, we have used the fact that $J_{1\over 2}$ is just
a sine function in disguise, then obsered that the integrand is manifestly even in $p$ and used this to rewrite the propagator as an integral from $(-\infty, \infty)$.
The graviton propagator can be similarly written as an integral over
the entire real line.
\be
\begin{split}
G^{\text{grav}}_{i j, k l} &=
\int_{-\infty}^{\infty} {-i dp \over \pi } \left[{\sin(p z) \sin (p z')   \over  z^2 (z')^2 \left(\vect{k}^2 + p^2 - i \epsilon\right)} \right.   \times {1 \over 2} \left.\left({\cal T}_{i k} {\cal T}_{j l} + {\cal T}_{i l} {\cal T}_{j k} - 
{2 {\cal T}_{i j} {\cal T}_{k l}\over d-1} \right)\right].
\end{split}
\ee

The advantage of writing the propagator as the third line of \eqref{axialpropagatorinf} is that when
we now obtain $p$-integrals in Witten diagrams these can be done 
just through an algebraic procedure of extracting residues. These simplifications happen for all odd boundary dimensions.
With this observation the recursion relations of \cite{Raju:2012zr}, specialized to $d=3$, can be written as:
\begin{equation}
\label{stressrecurs}
\begin{split}
&T(h_1, \vect{k_1}(w), \ldots h_4, \vect{k_4}(w)) =  \int_{-\infty}^{\infty} \left[ \sum_{\pi} {\cal I}_{\pi}(w,p) + {\cal B}(w,p) \right] d p  \\ 
&{\cal I}_{\pi}(w,p) = {p \over 2\cdot 2^s} \sum_{h^{\text{int}}, \pm} {-i {\cal T}^2 \over p^2 + (\vect{k_{\pi_1}}(w) + \vect{k_{\pi_2}}(w))^2}  {w-w^{\mp}(p) \over w^{\pm}(p) - w^{\mp}(p)}
\\
&{\cal T}^2 \equiv  T^*({h_{\pi_1}}, \vect{k_{\pi_1}}(p), {h_{\pi_2}}, \vect{k_{\pi_2}}(p), {h_{\text{int}}}, \vect{k_{\text{int}}}) T^*({-h_{\text{int}}}, -\vect{k_{\text{int}}},{h_{\pi_3}}, \vect{k_{\pi_3}}(p),  {h_{\pi_4}}, \vect{k_{\pi_4}}(p)).
\end{split}
\end{equation}
Here $T(\vect{h^1}, \vect{k^1}(w), \ldots \vect{h^4}, \vect{k^4}(w))$ is the
four point correlator with momenta $\vect{k^m}(w)$ extended according to \eqref{risagerextension} and 
polarization vectors that are specified in terms of the helicity by \eqref{polarizationvects}. The $T^*$ that appear on the right-hand sides are 
three point transition amplitudes that were computed in section \ref{secthreept}. 
As explained in \cite{Raju:2012zr}, ${\cal B}$ is a polynomial in $w$ with coefficients that
are rational functions of $p$; this term ensures that
the $p$-integral converges and also that the correlator to have the right behaviour at large $w$. We show below that we do not need to evaluate this term explicitly for the four point function. 

The reader should also note that we have a leading factor of ${1 \over 2 \cdot 2^s}$ in the definition of ${\cal I}_{\pi}$ compared to \cite{Raju:2012zr}. Here $s = 1$ for currents and $s = 2$ for the stress tensor. The factor of ${1 \over 2}$ comes from the fact that our integral runs over $(-\infty, \infty)$ instead of $(0,\infty)$. The second factor of ${1 \over 2^s}$ 
comes from the normalization of our polarization vectors in \eqref{normalizationpol}. 

We are actually interested in the original undeformed correlator, which is recovered by setting $w = 0$
in \eqref{stressrecurs}. We have:
\be
\label{stressrecursw0}
{\cal I}_{\pi}(0,p) =  {p \over 2 \cdot 2^s} \sum_{h_{\text{int}}, \pm} {i {\cal T}^2    \over p^2 + (\vect{k_{\pi_1}} + \vect{k_{\pi_2}})^2}  {w^{\mp}(p) \over w^{\pm}(p) - w^{\mp}(p)}.
\ee

\paragraph{Partitions:}
The recursion relations \eqref{stressrecurs} involve a sum
over partitions. For a non-color-ordered amplitude, we need to sum
over three partitions in the four point function. These three partitions
are
\be
\label{partset}
\pi \in \Big\{\{1,2,3,4\}, \{1,3,2,4\}, \{1,4,3,2\} \Big\}.
\ee
We will also call these partitions the $s$, $t$, and $u$ partitions respectively. 

\paragraph{$w^{\pm}$ as a function of $p$:}
To use the recursion relations we need to specify $w^{\pm}$ as a function of $p$. Consider a partition of the four external legs, described by $\pi$. Then
the pole under the extension \eqref{risagerextension} is at the value
of $w = w^{\pm}$ where
\be
\label{wlocation}
\begin{split}
&\left(\vect{k_{\pi_1}} + \alpha_{\pi_1} \vect{\ep_{\pi_1}} w^{\pm} + \vect{k_{\pi_2}} +  \alpha_{\pi_2} \vect{\ep_{\pi_2}} w^{\pm}\right)^2 + p^2 = 0. \\
\end{split}
\ee
We can write $w^{\pm} = u \pm v$, where we have defined the auxiliary quantities
\be
\begin{split}
&u  = -\frac{ \alpha_{\pi_1} \dotp[\ep_{\pi_{1}},k_{\pi_{2}}]+  \alpha_{\pi_2} \dotp[\ep_{\pi_{2}}, k_{\pi_{1}}]} {2  \alpha_{\pi_1}  \alpha_{\pi_2} \dotp[\ep_{\pi_{1}},\ep_{\pi_{2}}]}, \\
&v = \frac{\sqrt{( \alpha_{\pi_1} \dotp[\ep_{\pi_{1}},k_{\pi_{2}}]+  \alpha_{\pi_2} \dotp[\ep_{\pi_{2}}, k_{\pi_{1}}])^2 -4  \alpha_{\pi_1}  \alpha_{\pi_2} \dotp[\ep_{\pi_{1}}, \ep_{\pi_{2}}]
   \left(p^2+(\vect{k_{\pi_{1}}}+ \vect{k_{\pi_{2}}})^2 \right)}}{2  \alpha_{\pi_1}  \alpha_{\pi_2} \dotp[\ep_{\pi_{1}},\ep_{\pi_{2}}]}.
\end{split}
\ee
As we will see below, we need to evaluate these expressions only at 
specific values of $p$ and in those cases, they often simplify considerably.

We should also stress that what is important is that there are two solutions for $w$, given a value of $p$. Which solution we call $w^+$ and which we call $w^{-}$ is of no relevance and we will be somewhat cavalier about this below. 

\paragraph{Rational Integrands:}
The integrands ${\cal I}_{\pi}(w,p)$ in \eqref{stressrecurs} might seem to have square-roots 
but, 
in fact, all these square roots cancel. This is because of the 
sum in front, which takes $v \leftrightarrow -v$. 
As a consequence, all the integrands depend only on even powers of
$v$, which means that there are no square-roots after accounting
for both terms. 

\paragraph{Intermediate Spinors:}
Now, let us consider the spinors for the intermediate leg. The intermediate momentum is just
\be
\label{intermom}
\vect{k}_{\text{int}} = -\vect{k_{\pi_1}}(w^{\pm}) - \vect{k_{\pi_2}}(w^{\pm}) = \vect{k_{\pi_3}}(w^{\pm}) + \vect{k_{\pi_4}}(w^{\pm}).   
\ee
The three point amplitudes above are written in terms of spinors. However, the key point is that in choosing a
decomposition of  $\vect{k}_{\text{int}}$ into spinors, we can rescale $ \la_{\rm int} \rightarrow \alpha \la_{\rm int}$
and $\lb_{\rm int} \rightarrow \alpha^{-1} \lb_{\rm int}$ by any complex number $\alpha$ without affecting the final answer. This is because
in the recursion relations above when we have $h_{\rm int}$ on the left, we have $-h_{\rm int}$ on the right and so $\la_{\rm int}$ and $\lb_{\rm int}$ always come together.  Consequently, we can choose the intermediate spinors using 
\eqref{lint}, and avoid any square roots.

Other, more covariant looking, choices are possible. For example one could take
\begin{equation}
\label{interspin}
\begin{split}
\la_{\rm int} &= \la_{\pi_{2}}(w^{\pm}) + \la_{\pi_{1}}(w^{\pm}) {i \left(\norm{k_{\pi_{1}}} - \norm{k_{\pi_{2}}} - i p \right) \over \dotlm[\lb_{\pi_{2}}(w^{\pm}), \la_{\pi_{1}}(w^{\pm})]} \\
\lb_{\rm int} &=  - \lb_{\pi_{2}}(w^{\pm}) - {i \left(\norm{k_{\pi_{1}}} + \norm{k_{\pi_{2}}} - i p \right) \over \dotl[\la_{\pi_{1}}(w^{\pm}), \la_{\pi_{2}}(w^{\pm})]}  \lad_{\pi_{1}}(w^{\pm}). \\
\end{split}
\end{equation}

In fact in the calculations of sections \ref{secmhvym} and \ref{secmhvgrav}, we will never need the intermediate spinors explicitly. Instead we will use identities like \eqref{isintspin} to rewrite expressions that involve $\la_{\rm int}$ like \eqref{ipi1prelim} as expressions that are free of these terms like \eqref{ipi1}.
 
\paragraph{Boundary Term ${\cal B}$:}
Now, we have argued above that the integrand term in \eqref{stressrecurs} is rational
and even in $p$. So, merely by polynomial division, we can write it in the following form:
\be
\sum_{\pi} {\cal I}_{\pi}(w,p) 
= {{\cal N}(p^2,w) \over {\cal D}(p^2,w)} + {\cal Q}(p^2, w),
\ee
where ${\cal N}, {\cal D}, {\cal Q}$ are polynomials and ${{\cal N} \over {\cal D}}$ dies off
at least as fast as ${1 \over p^2}$ for large $p$. The fact that ${\cal Q}$ is polynomial in $w$ follows from the fact that the highest power of $p$ in ${\cal D}$ is independent of $w$. (Note that ${\cal T}^2$ is purely a function of $p$ and the only dependence on $w$ in ${\cal I}_{\pi}(w,p)$ comes through the factors that are explicitly displayed.)

To ensure the convergence of the integral at large $p$ and the correct behaviour of the correlator at large $w$, we need to set:
\be
{\cal B}(p, w) = -{\cal Q}(p^2, w) + \sum_m b_m(p) w^m,
\ee
where $b_m(p)$ are rational functions of $p$ with a convergent integral over the real line. For conserved currents, we can take $b_m(p) = 0$ since the correlator vanishes at large $w$. 

For stress tensor correlators, the behaviour of the correlator at large $w$ is completely fixed by the Ward identities as shown in \cite{Raju:2012zr} and we can take the $b_m(p)$ to be any functions that satisfy:
\be
T(h_1, \vect{k_1}(w), \ldots h_4, \vect{k_4}(w)) \underset{w \rightarrow \infty}{\longrightarrow}\sum_m w^m \int_{-\infty}^{\infty} b_m(p) d p.
\ee
However, we know that the Ward identities contribute only {\em local terms} at large $w$; in momentum space, this corresponds to terms that are analytic in at least two momenta. These terms 
are not themselves of physical interest and so, at the level of the four point function, we can just forget about the $b_m$ functions.
\paragraph{Algebraic Evaluation of the $p$ Integral:}
We now show that the 
entire $p$ integral can be done just by picking out residues of the integrand at pre-specified
poles.  Now that we have dealt with the behaviour of the integrand at large $w$, we will specialize to $w = 0$ for simplicity. 

Since, with the addition of ${\cal B}$ the integrand vanishes at large $p$ by construction, we
can close the contour through either the upper or the lower half plane. This leaves us just
with the task of evaluating some residues. In fact we do not need to evaluate ${\cal Q}$ explicitly either:
\be
\int_{-\infty}^{\infty}  {{\cal N}(p^2, w) \over {\cal D}(p^2, w)} d p = 2 \pi i \,  \sum_{\text{poles}} {\text{Res}} \left[I_{\pi}(w,p) \right], 
\ee
where we sum over the poles of the integrand at finite $p$ in the upper half plane. 

Second, it is, in fact, quite easy to specify the locations of {\em all poles} in the integrand.
There are two sources of poles: (a) the poles in the three point function where $p = \pm i (\norm{k_{\pi_1}} + \norm{k_{\pi_2}})$ (b) the pole where the propagator vanishes $p^2 + (\vect{k_{\pi_1}} + \vect{k_{\pi_2}})^2 = 0$.

So, for each partition $\pi$,  there are exactly {\em three poles} in the upper half plane. This
set is given by:
\be
\label{polelist}
{\cal P}_{\pi} = \left\{i (\norm{k_{\pi_1}} +  \norm{k_{\pi_2}}), i (\norm{k_{\pi_3}} +  \norm{k_{\pi_4}}), i \sqrt{\big(\vect{k_{\pi_1}} + \vect{k_{\pi_2}}\big)^2} \right\}.
\ee

This leads to our final formula for the four point function:
\be
\label{formulafinal}
\boxed{T(h^1, \vect{k^1}, \ldots h^4, \vect{k^4}) = 2 \pi i \sum_{\pi} \sum_{p_0 \in {\cal P}_{\pi}} \underset{p = p_0}{\text{Res}} \left[{\cal I}_{\pi}(0,p) \right],
}
\ee
Here ${\cal I}(0,p)$ is specified in \eqref{stressrecursw0}. The three point transition amplitudes
that appear there are specified in section \ref{secthreept}, with intermediate momenta and spinors given by \eqref{intermom} and \eqref{interspin}. Moreover, $w^{\pm}(p)$ is specified by \eqref{wlocation}, the set of partitions $\pi$ is specified by \eqref{partset}, and the set of poles ${\cal P}$ is specified by \eqref{polelist}. 

This leads to a straightforward algorithm that is implemented in the attached Mathematica code. We also evaluate this formula in several cases below. 

\paragraph{An Aside:}
Before we conclude this section, let us comment briefly on the various kinds of terms that appear in the formula above. First, note that the residue at $p = i (\norm{k_{\pi_1}} +  \norm{k_{\pi_2}})$ or $p =  i (\norm{k_{\pi_3}} +  \norm{k_{\pi_4}})$ is a rational function of the external spinors.  This is guaranteed since $I_{\pi}$ is a rational function and so is the location of the pole. Furthermore, the analysis of \cite{Raju:2012zr} (and our explicit computations below) tells us that when we take the flat space limit, it is these two terms that give us the correct singularity in the final answer.

On the other hand the last entry
in the set \eqref{polelist} is not important in the flat space limit. It also has a different analytic structure, and it can be written as a rational function of the external spinors and the norms of the sums of momenta. If we choose to write it purely as a function of the original spinors then we get square roots in this term, which arise because the location of the pole involves a square root. 

However, this term has an interesting relation to the operator product expansion that we should mention. First,  it is easy to check that for this pole one of the possible solutions for $w^{\pm}(p)$ is just  $w^{-} = 0$. (There is another solution to \eqref{wlocation} but since one solution is $0$ this does not contribute due to the ${w^{\mp} \over w^{pm} - w^{\mp}}$ factor in front of the integrand.)  The residue at this pole is merely:
\[
 2 \pi i \sum_{h_{\text{int}}} T^*\big(h_1, \vect{k_{\pi_1}}, h_2, \vect{k_{\pi_2}}, h_{\text{int}}, \vect{k_{\pi_1}} + \vect{k_{\pi_2}} \big)  T^*\big(h_1, \vect{k_{\pi_1}}, h_2, \vect{k_{\pi_2}}, -h_{\text{int}}, \vect{k_{\pi_1}} + \vect{k_{\pi_2}} \big) {i \over 4}
\]
Now the correlator is obtained by contracting the bulk vertices
with a bulk-boundary propagator. From the relations between Bessel functions:
\be
z^{\nu} J_{\nu}(p z) = {- 2 i \over \pi p^{\nu}} (-i p z)^{\nu} K_{\nu}(-i p z) - i z^{\nu} N_{\nu}(p z),
\ee
it is easy to check that one term in the transition amplitude is the correlator:
\be
\label{transitioncorrelation}
T^*\big(h_1, \vect{k_{\pi_1}}, h_2, \vect{k_{\pi_2}}, h_{\text{int}}, \vect{k_{\pi_1}} + \vect{k_{\pi_2}} \big) = -i \sqrt{2 \over \pi} {T\big(h_1, \vect{k_{\pi_1}}, h_2, \vect{k_{\pi_2}}, h_{\text{int}}, \vect{k_{\pi_1}} + \vect{k_{\pi_2}} \big) \over \big(i \norm{k_{\pi_1} + k_{\pi_2}}\big)^\nu} + \ldots
\ee
where $\nu = {1 \over 2}$ for currents and $\nu = {3 \over 2}$ for the stress tensor and where $\ldots$ is the term that comes from the Neumann function above. So we see that one term in our final answer is exactly the product of the
three point functions divided by the two point function as predicted by the operator product expansion.

We should emphasize that although this conformal block drops out in the flat space limit, the Witten diagram involving the exchange of a graviton does not. The flat space limit of \cite{Raju:2012zr} was derived diagram by diagram; so the exchange Witten diagram goes over to the flat space exchange diagram. This is consistent because the exchange Witten diagram involves more than just the conformal block of the stress-tensor (as is discussed, for example, in section 6.4 of \cite{ElShowk:2011ag}) and so it survives in the flat space limit.

Here should caution the reader that although the pole at $p = i |\vect{k_{\pi_1}} + \vect{k_{\pi_2}}|$ accounts for the contribution of the conformal block of the stress tensor or the conserved current itself, we have not shown that the remainder of the correlator including the $\ldots$ in \eqref{transitioncorrelation} and the contribution from the other poles can be exactly accounted for by the contribution of all double trace operators.\footnote{There is, of course, an infinite sequence of such double trace operators whose exact spectrum can be easily worked out using character decomposition \cite{Barabanschikov:2005ri}. For example, below conformal weight 8, we find the spectrum of double trace operators of the stress tensor (in the notation [weight, spin]): $(6,0) \oplus (6,1) \oplus (6,2) \oplus (6,3) \oplus (6,4) \oplus (7,0) \oplus (7,1) \oplus (7,2) \oplus (7,3) \oplus 2 (7,4) \oplus (7,5)$.}
 This deserves some further study.

\section{MHV Correlators for Conserved Currents \label{secmhvym}}
In this section, we will expand the formula above in terms of spinors for the 
four point function of currents. We start by analyzing the color-ordered MHV correlator $+-+-$, and then describe the full (i.e. non-color-ordered) MHV correlator.  Although the calculations below might seem tedious, our final answer
for the color ordered MHV correlator is quite simple and is given in \eqref{colororderedmhv}. The expressions below are also implemented in the attached Mathematica program (available from the source file in the arXiv submission), which may be useful while following this analysis.  We show, explicitly, for both the 
color ordered and full MHV amplitude that taking the flat space limit just
leads to the Parke-Taylor formula for four-gluon scattering. 

\subsection{Color-Ordered MHV Correlator}

To obtain the color ordered amplitude we only need to sum over two partitions: the (12)(34) partition and the (41)(23) partition. 

Let us start by analyzing the (12)(34) partition, which we will call
the ``s'' partition. In fact the ``t'' partition: (14)(23) is just obtained
by taking all the results here and interchanging $2 \leftrightarrow 4$. So that will not require a separate calculation. 

Let us expand out the three point functions that
appear in the formula above. Doing this, we find:
\be
\label{ipi1prelim}
\begin{split}
{\cal I}_{s} = \sum_{\pm} &\Bigg\{{ R^{\text{YM}}(|\vect{k_1}|, |\vect{k_2}|, p) \over 2 \sqrt{2}  |\vect{k_1}| |\vect{k_2}| p} \bigl(\norm{k_2} + i p  - \norm{k_1} \bigr) \bigl(i p + \norm{k_1} - \norm{k_2} \bigr) \bigl(\norm{k_1} + \norm{k_2} - i p \bigr) \\
&{ R^{\text{YM}}(|\vect{k_3}|, |\vect{k_4}|, -p) \over 2 \sqrt{2}  |\vect{k_3}| |\vect{k_4}| p} \bigl(\norm{k_4} - i p  - \norm{k_3} \bigr) \bigl(-i p + \norm{k_3} - \norm{k_4} \bigr) \bigl(\norm{k_3} + \norm{k_4} + i p \bigr) \\
&\begin{split} \times \Bigg[ &\left({\dotlb[\lb_{1}, \lb_{\text{int}}]^3 \over \dotlb[\lb_{1}, \lb_{2}(w^{\pm})] \dotlb[\lb_{2}(w^{\pm}), \lb_{\text{int}}] } {\dotl[\la_{4}, \la_{\text{int}}]^3 \over \dotlb[\la_{3}(w^{\pm}), \la_{4}] \dotlb[\la_{3}(w^{\pm}), \la_{\text{int}}] }\right) \\ &+ \left({\dotl[\la_2, \la_{\text{int}}]^3 \over \dotl[\la_{1}(w^{\pm}), \la_{2}] \dotl[\la_{1}(w^{\pm}), \la_{\text{int}}] } {\dotlb[\lb_{3}, \lb_{\text{int}}]^3 \over \dotlb[\lb_{3}, \lb_{4}(w^{\pm})] \dotlb[\lb_{4}(w^{\pm}), \lb_{\text{int}}] }\right)
\Bigg]  \Bigg\}
\end{split}\\
&\times {1 \over 4} {i p \over \left(p^2 + (\vect{k_1} + \vect{k_2})^2\right)} {w^{\mp} \over w^{\pm} - w^{\mp}}
\end{split}
\ee
We can simplify by expanding out $R^{\text{YM}}$ and also recognizing
that
\be
\label{isintspin}
\begin{split}
& \la_1(w^{\pm}) \lb_1 + \la_2 
\lb_2 (w^{\pm})+ \la_{\rm int} \lb_{\rm int}  = i (\norm{k_1} + \norm{k_2} + i p) \sigma^3. \\
\end{split}
\ee
This leads to 
\be
\label{ipi1}
\begin{split}
&{\cal I}_s  = \sum_{\pm}  
\Bigg\{
\frac{
 \bigl(\norm{k_2} + i p  - \norm{k_1} \bigr) \bigl(i p + \norm{k_1} - \norm{k_2} \bigr)  \bigl(\norm{k_3} + i p  - \norm{k_4} \bigr) \bigl(i p + \norm{k_4} - \norm{k_3} \bigr) }{16 \pi |\vect{k_{1}}|  |\vect{k_{2}}| |\vect{k_{3}}|  |\vect{k_{4}}|  \left(|\vect{k_{1}}|+|\vect{k_{2}}|+ i p \right)  \left(|\vect{k_{3}}|+|\vect{k_{4}}|- i p \right)} \\ 
&\begin{split} 
\times \Bigg[&{ 
\Big(
\dotlb[\lb_1, \lb_2(w^{\pm})] 
\dotl[\la_{4}, \la_2] + i E_p^{12} 
\dotlm[\lb_1,\la_4] \Big)^3
  \over \dotlb[\lb_{1}, \lb_{2}(w^{\pm})] \dotlb[\la_4,\la_{3}(w^{\pm})]
\left(\dotlb[\lb_{2}(w^{\pm}), \lb_{1}] \dotl[\la_1(w^{\pm}), \la_3(w^{\pm})] - i E_p^{12} \dotlm[\la_3(w^{\pm}), \lb_2(w^{\pm})] \right)  }
\\ &
+{\left(\dotl[\la_2, \la_1(w^{\pm})]\dotlb[\lb_{3}, \lb_{1}]
+ i E_p^{12} \dotlm[\la_2, \lb_3] \right)^3  \over 
\dotl[\la_{1}(w^{\pm}), \la_{2}] \left(\dotl[\la_{1}(w^{\pm}), \la_2] \dotlb[\lb_{4}(w^{\pm}), \lb_{2}(w^{\pm})] + i E_p^{12} \dotlm[\la_1(w^{\pm}), \lb_4(w^{\pm})] \right) \dotlb[\lb_{3}, \lb_{4}(w^{\pm})] }
\Bigg]
\end{split}
\\ &\times {-i \over p^2 + (\vect{k_1} + \vect{k_2})^2} {w^{\mp} \over w^{\pm} - w{\mp}}  \Bigg\}.
\end{split}
\ee
where we have defined
\be
E_p^{n m} \equiv i p + \norm{k_n} + \norm{k_m}.
\ee

As we described in section \ref{secformula}, it is clear that there are two kinds of poles 
that appear in ${\cal I}_{s}$. One type is the pole that appears
from the constituent three point amplitudes: the existence of such a pole
is required by the fact that the three point amplitude must have the correct flat space limit i.e. the flat space three point amplitude must appear as the coefficient of singularities at $i p + \norm{k_{1}} + \norm{k_{2}} = 0$ and $i p + \norm{k_{3}} + \norm{k_{4}} = 0$.  The second kind of pole appears
when the propagator factor above $p^2 + (\vect{k_{1}} + \vect{k_{2}})^2$ 
vanishes. 

\paragraph{Poles from the three point amplitude: }
We have written the expression above so that it is very easy to extract the
pole at $E_p^{12} = 0$.  First, let us note that when $E_p^{12} = 0$,
the value of $w^{\pm}(p)$ simplifies. Denoting this value by $w^{\pm}(p = i(\norm{k_1} + \norm{k_2})) \equiv w_{s_1}^{\pm}$, we have
\begin{equation}
\Bigl( (\la_1  + \beta_1 \lbd_1 w^{\pm}_{s_1}) \lb_1 + \la_2 (\lb_2 + \beta_2 \lad_2 w^{\pm}_{s_1}) \Bigr)^2 = 0.
\end{equation}
This condition requires either 
\begin{equation}
\dotl[\la_1 + \beta_1 \lbd_1 w^{\pm}_{s_1}, \la_2] = 0,
\quad \text{or} \quad 
\dotl[\lb_1, \lb_2 + \beta_2 \lad_2 w^{\pm}_{s_1}] = 0.
\end{equation}
These equations are solved by
\begin{equation}
\label{ws1p}
\beta_1 w^+_{s_1} = - {\dotl[\la_2, \la_1] \over \dotlm[\la_2, \lb_1]};
\end{equation}
or 
\begin{equation}
\label{ws1m} 
\beta_2 w^-_{s_1} = -{\dotlb[\lb_1, \lb_2] \over \dotlm[\lb_1, \la_2]}.
\end{equation}
We remind the reader that the $\beta_m$ are given by \eqref{explicitbeta}.

To proceed further we recognize that at the pole $w_{s_1}^+$ the 
second line of the big square bracket in \eqref{ipi1}, which corresponds
to $h_{\text{int}} = -1$, vanishes. So we only need to evaluate the first line in the big square bracket at the point where $E_p^{12} = 0$. 

Some short calculations tell us that, the propagator factor simplifies
\be
\label{propsimplew}
\begin{split}
{ i \over -(\norm{k_1} + \norm{k_2})^2 + (\vect{k_1} + \vect{k_2})} {w_{s_1}^{\mp} \over w_{s_1}^{\pm} - w_{s_1}^{\mp}} &= {i \over \dotl[\la_1, \la_1] \dotl[\lb_1, \lb_2]} {w_{s_1}^{\mp} \over w_{s_1}^{\pm} - w_{s_1}^{\mp}}  \\ &= {1 \over \dotlb[\lb_1, \lb_2(w_{s_1}^+)] \dotl[\la_1, \la_2]}.
\end{split}
\ee
Another short calculation tells us that
\be
\label{explicitspinsws1p}
\begin{split}
\la_1(w_{s_1}^+) &= \la_2  {\dotlm[\la_1,\lb_1] \over \dotlm[\la_2,\lb_1]},\\
\la_3(w_{s_1}^+) &= {\dotlm[\la_3,\lb_3] + \dotlm[\la_4,\lb_4] - i E^T\over \dotlm[\la_2,\lb_3]} \la_2 - \la_4 {\dotlm[\la_4,\lb_4] - i E^T \over \dotlm[\la_4,\lb_3]}. 
\end{split}
\ee

Plugging these factors in, we find that
\begin{align}
\label{firstexp1}
(2 \pi i) \underset{p = i (\norm{k_1} + \norm{k_2})}{\text{Res}} \big[I_{s}(p) \big] &= 
{E^{124,3} \over 4 \norm{k_3} \norm{k_4} \norm{k_1}} {\dotl[\la_2, \la_4] \dotlm[\la_4, \lb_3] \dotlm[\la_2, \lb_3] \dotlm[\la_2,\lb_1]  \over E^{12, 34} \dotl[\la_1, \la_2] E^T}
\\ 
&\label{secondexp1} + (\la_1, \la_2, \la_3, \la_4) \leftrightarrow (\lb_2, \lb_1, \lb_4, \lb_3).
\end{align}
Here in \eqref{secondexp1} we have recognized that the contribution from \eqref{ws1m} is just obtained by the interchange indicated. 

\paragraph{Poles from the propagator:}
We now turn to the second kind of pole, which comes when the propagator vanishes.  When $p^2 + (\vect{k_1} + \vect{k_2})^2 = 0$, one of the solutions --- which we will denote by $w^{-}\Big(\sqrt{ (\vect{k_1} + \vect{k_2})^2} \Big)$ --- is just 0. The other solution is complicated, but it does not contribute to the answer at all; the factor ${w^{-} \over w^{+} - w^{-}}$ that appears in \eqref{ipi1} vanishes since $w^{-} = 0$ at this value of $p$. 
\begin{align}
\label{aschannel1} {\cal A}_s = (2 \pi i) \underset{p = i \sqrt{(\vect{k_1} + \vect{k_2})^2}}{\text{Res}}&\Big[{\cal I}_{s}(p) \Big] = { i E^{1,2 s} E^{2,1 s} E^{4,3 s} E^{3,4 s} \over 16 E^{12,s} E^{34s} \norm{k_1 + k_2}} \\ \label{aschannel2} &\times \Bigg[{\left( \dotl[\la_2, \la_1] \dotl[\lb_3, \lb_1] + i E^{12,s} \dotlm[\la_2, \lb_3] \right)^3 \over \dotl[\la_1, \la_2] \dotlm[\lb_3, \lb_4] \left(\dotl[\la_1, \la_2] \dotl[\lb_4, \lb_2] + i E^{12,s} \dotl[\la_1, \lb_4]\right) }
\\ \label{aschannel3} & +{( \dotl[\la_4, \la_2] \dotl[\lb_1, \lb_2] + i E^{12,s} \dotlm[\la_4, \lb_1])^3 \over \dotl[\la_4, \la_3] \dotl[\lb_1, \lb_2] (\dotl[\la_1, \la_3] \dotl[\lb_2, \lb_1] - i E^{12,s} \dotlm[\la_3, \lb_2])}
 \Bigg].
\end{align}

As we mentioned above,this residue is quite interesting since it contributes exactly the product of the undeformed transition amplitude on the left and the right. This contains the contribution of the conformal block of the current itself and is consistent with what we would expect from the operator product expansion applied
in momentum space. However, we repeat the caveat that it is necessary to 
also show that the remaining terms are consistent with the contribution of double trace operators.

\subsubsection{Final answer for the color ordered current correlator}
The final answer for the color ordered MHV current correlator can now  just be obtained by interchanges from the 
answer above. 
The answer for the four point amplitude is given by
\begin{equation}
\label{colororderedmhv}
T^{+-+-}(\vect{k_1}, \vect{k_2}, \vect{k_3},\vect{k_4}) =  {{\cal F} \over E^T} + {\cal A},
\end{equation}
where ${\cal F}$ is the term with a pole that corresponds to the flat space limit and ${\cal A}$ is an intrinsically AdS term that
would vanish in flat space. We have,
\begin{align}
\label{firstexp} {\cal F} &= 
{E^{124,3} \over 4 \norm{k_3} \norm{k_4} \norm{k_1}} {\dotl[\la_2, \la_4] \dotlm[\la_4, \lb_3] \dotlm[\la_2, \lb_3] \dotlm[\la_2,\lb_1]  \over E^{12, 34} \dotl[\la_1, \la_2]}
\\ \label{secondexp} &+ (\la_1, \la_2, \la_3, \la_4) \leftrightarrow (\lb_2, \lb_1, \lb_4, \lb_3) \\ &+  (\la_1, \la_2, \la_3, \la_4, \lb_1, \lb_2, \lb_3, \lb_4) \leftrightarrow  ( \la_3, \la_4, \la_1, \la_2, \lb_3, \lb_4, \lb_1, \lb_2) \\ \label{thirdexp} &+  (\la_1, \la_2, \la_3, \la_4, \lb_1, \lb_2, \lb_3, \lb_4) \leftrightarrow  ( \lb_4, \lb_3, \lb_2, \lb_1, \la_4, \la_3, \la_2, \la_1) \Bigg]
\\ \label{tchannel} &+ (\la_2, \lb_2) \leftrightarrow (\la_4, \lb_4). 
\end{align}
Note that within the big square bracket, all interchanges correspond to the expression in \eqref{firstexp}. So, for example to get
\eqref{thirdexp} we take \eqref{firstexp} and perform the interchanges indicated. However to get \eqref{tchannel} we take the whole
square bracket and perform the interchange indicated.
For ${\cal A}$ we have
\be
{\cal A} = {\cal A}_s + (\la_2, \lb_2) \leftrightarrow (\la_4, \lb_4),
\ee
where ${\cal A}_s$ is specified in 
the three lines \eqref{aschannel1}, \eqref{aschannel2}, and \eqref{aschannel3}. We take all three lines and perform the substitution indicated.

\subsubsection{Flat Space Limit of the Answer \label{subsecflatspacecurr}}
We can, quite easily, take the flat space limit of the answer above. We just need to look at the ${\cal F}$ term above. Second, at $E^T=0$, various spinor identities can be used to simplify the function.

However, here we will take a different route that is somewhat more elegant, and also gives us a check on the final answer.  Consider the functions 
\begin{equation}
\label{mhvbarmhv}
\begin{split}
&M^{\text{MHV}}(w) = {\dotl[\la_2, \la_4]^4 \over \dotl[\la_1(w), \la_2] \dotl[\la_2, \la_3(w)] \dotl[\la_3(w), \la_4] \dotl[\la_4, \la_1(w)]}, \\ 
 &M^{\overline{\text{MHV}}}(w) = {\dotl[\lb_1, \lb_3]^4 \over \dotl[\lb_1, \lb_2(w)] \dotl[\lb_2(w), \lb_3] \dotl[\lb_3, \lb_4(w)] \dotl[\lb_4(w), \lb_1]},
\end{split}
\end{equation}
where the spinors are extended as above. Note that, as above, $\la_2, \la_4, \lb_1, \lb_4$ do not change in this extension. 

Then \eqref{firstexp} is actually just proportional to the residue of $M^{\text{MHV}}(w)$ at $w_{s_1}^+$, while \eqref{secondexp} is 
just proportional to the residue of  $M^{\overline{\text{MHV}}}(w)$ at $w_{s_1}^-$ In fact, we have
\begin{equation}
\label{residuesflatspace}
\begin{split}
&\eqref{firstexp} =  {E^{123,4} E^{124,3} \over 2 \norm{k_3} \norm{k_4} E^T} \lim_{w \rightarrow w_{s_1}^+} {\dotl[\la_1(w), \la_2] \over \dotl[\la_1(0), \la_2]} M^{\text{MHV}}(w),\\
&\eqref{secondexp} =  {E^{123,4} E^{124,3} \over 2 \norm{k_3} \norm{k_4} E^T} \lim_{w \rightarrow w_{s_1}^-} {\dotl[\lb_1, \lb_2(w)] \over \dotl[\lb_1, \lb_2(0)]} M^{\overline{\text{MHV}}}(w).
\end{split}
\end{equation}
These are true as {\em exact} statements without setting $E^T = 0$. However, when we set $E^T = 0$, the factor in front of the limit just becomes $1$.  Then we can see that by adding together the various terms in the expression for ${\cal F}$, we will just get the sum of 
the holomorphic and anti-holomorphic MHV amplitudes as the residue of the pole at $E^T = 0$. These amplitudes are, of course, equal in the flat space limit. 

To prove the assertion \eqref{residuesflatspace}, consider the expression for the integrand \eqref{ipi1prelim}. 
At the pole $w_{s_1}^+$, we have $\la_1(w_{s_1}^+) \propto \la_2$. So, the spinor expressions that appear in \eqref{ipi1prelim} are
\be
\label{ws1spinor}
{\dotl[\lb_1, \lb_{\rm int}]^3 \over \dotl[\lb_1, \lb_2(w_{s_1}^+)] \dotl[\lb_2(w_{s_1}^+), \lb_{\rm int}]} {\dotl[\la_4, \la_{\rm int}]^3 \over \dotl[\la_3(w_{s_1}^+), \la_4] \dotl[\la_{\rm int}, \la_3(w_{s_1}^+)]} {1 \over \dotl[\la_1, \la_2] \dotlb[\lb_1, \lb_2(w_{s_1}^+)]},
\ee
where the third term comes from the propagator after using \eqref{propsimplew}.

However, we can write
\be
 \dotl[\lb_1, \lb_{\rm int}] \dotl[\la_4, \la_{\rm int}] = -\dotl[\lb_1, \lb_2(w_{s_1}^+)] \dotl[\la_4, \la_2],
\ee
where we have just used the fact that 
\be
-\la_{\rm int} \lb_{\rm int} = \la_1 (w_{s_1}^+) \lb_1 + \la_2 \lb_2 (w_{s_1}^+).
\ee
Using this to simplify both the numerator and the denominator we find that the spinor expression becomes 
\be
\begin{split}
\eqref{ws1spinor}  &= 
{\dotl[\lb_1, \lb_2(w_{s_1}^+)]^3 \over \dotl[\lb_1, \lb_2(w_{s_1}^+)] \dotl[\lb_2(w_{s_1}^+), \lb_1]} {\dotl[\la_4, \la_{2}]^3 \over \dotl[\la_3(w_{s_1}^+), \la_4] \dotl[\la_1(w_{s_1}^+), \la_3(w_{s_1}^+)]} {1 \over \dotl[\la_1, \la_2] \dotlb[\lb_1, \lb_2(w_{s_1}^+)]} \\
&= {\dotl[\la_4, \la_2]^3 \over \dotl[\la_1, \la_2] \dotl[\la_1(w_{s_1}^+),  \la_3(w_{s_1}^+)] \dotl[\la_3(w_{s_1}^+), \la_4]} \\
&= {\dotl[\la_4, \la_2]^4 \over \dotl[\la_1, \la_2] \dotl[\la_2, \la_3(w_{s_1}^+)] \dotl[\la_2,  \la_3(w_{s_1}^+)] \dotl[\la_4, \la_1(w_{s_1}^+)]},
\end{split}
\ee
where in the last step we have used the identity that
\be
\dotl[\la_2, \la_4] \dotl[\la_1(w_{s_1}^+), \la_3(w_{s_1}^+)] = \dotl[\la_1(w_{s_1}^+), \la_4] \dotl[\la_2, \la_3(w_{s_1}^+)].
\ee
Equation \eqref{residuesflatspace} now follows immediately when we recognize the pre-factor that appears in the correlator. 

We can now use \eqref{residuesflatspace} to independently reconstruct an expression for the amplitude. This is because
given the function $M(w)$ defined in \eqref{mhvbarmhv}, we can force it have the correct residues at the four poles
$w_{s_1}^+, w_{s_2}^+, w_{t_1}^+, w_{t_2}^+$. The simplest way to do this is by Lagrange interpolation. 

We construct the rational function
\be
\label{madsdef}
\begin{split}
M^{\text{AdS}}(w) &= M^{\text{MHV}}(w) P^{+}(w) \\
M^{\overline{\text{AdS}}}(w) &= M^{\overline{\text{MHV}}}(w) P^{-}(w) \\
P^{\pm}(w) &\equiv  
{(w - w_{s_2}^{\pm}) (w - w_{t_1}^{\pm}) (w - w_{t_2}^{\pm}) \over (w_{s_1}^{\pm} - w_{s_2}^{\pm}) (w_{s_1}^{\pm} - w_{t_1}^{\pm}) (w_{s_1}^{\pm} - w_{t_2}^{\pm})} {E^{123,4} E^{124,3} \over 2 \norm{k_3} \norm{k_4}}
\\ &+  {(w - w_{s_1}^{\pm}) (w - w_{t_1}^{\pm}) (w - w_{t_2}^{\pm}) \over (w_{s_2}^{\pm} - w_{s_1}^{\pm}) (w_{s_2}^{\pm} - w_{t_1}^{\pm}) (w_{s_2}^{\pm} - w_{t_2}^{\pm})} {E^{134,2} E^{234,1} \over 2 \norm{k_1} \norm{k_2}} \\ &+ 
{(w - w_{s_1}^{\pm}) (w - w_{s_2}^{\pm}) (w - w_{t_1}^{\pm}) \over (w_{t_1}^{\pm} - w_{s_2}^{\pm}) (w_{t_1}^{\pm} - w_{s_2}^{\pm}) (w_{t_1}^{\pm} - w_{t_2}^{\pm})} {E^{134,2} E^{124,3} \over 2 \norm{k_3} \norm{k_2}} \\ &+  
{(w - w_{s_1}^{\pm}) (w - w_{s_2}^{\pm}) (w - w_{t_2}^{\pm}) \over (w_{t_2}^{\pm} - w_{s_1}^{\pm}) (w_{t_2}^{\pm} - w_{s_2}^{\pm}) (w_{t_2}^{\pm} - w_{t_1}^{\pm})} {E^{134,2} E^{234,1} \over 2 \norm{k_1} \norm{k_2}}.
\end{split}
\ee
$P^+(w)$ is a Lagrange polynomial with the property that modulates the residues of $M^{\text{MHV}}(w)$ to produce the correct residues
required in $M^{\text{AdS}}(w)$. Note that $M^{\text{AdS}}(w)$ still has the desired falloff at infinity because $M^{\text{MHV}}(w) \rightarrow {1 \over w^4}$ at large $w$ and $P^+(w) \rightarrow w^3$ at large $w$.

In terms of these functions, we have
\be
\label{fintermsofads}
{\cal F} = M^{\text{AdS}}(0) + M^{\overline{\text{AdS}}}(0).
\ee
In the flat space limit, we have the {\em identity}
\be
P^{+}(w) = P^{-}(w) = 2, \quad \text{at} ~E^T = 0.
\ee
So, it is clear that in the flat space limit, we have
\be
T^{+-+-} \underset{E^T \rightarrow 0}{\longrightarrow} {4 \over E^T} M^{\text{MHV}} = {4 \over E^T} M^{\overline{\text{MHV}}},
\ee
which is exactly what we expect from the flat space conjecture.\footnote{The extra factor of $4$ comes, once again, from the unconventional normalization of our polarization vectors.}

Finally we notice that this method provides a check on our answer in \eqref{firstexp}. 
We can now do some algebra given the explicit positions of the poles. We have already computed the values of $w_{s_1}^{\pm}$ in \eqref{ws1p} and all other poles are given by obvious substitutions in those equations. For example, we have
\begin{align}
\label{ws2p}
&\beta_3 w_{s 2}^+ = - {\dotl[\la_4, \la_3] \over \dotlm[\la_4,\lb_3]},\\
\label{ws2m}
&\beta_4 w_{s 2}^- = -{\dotlb[\lb_3, \lb_4] \over \dotlm[\lb_3,\la_4]}.
\end{align}
We see that
\be
\label{pwfact1}
\begin{split}
{1 \over {w_{s_1}^+ \over w_{s_2}^+} - 1} &= {-1 \over {\dotl[\la_2, \la_1] \dotlm[\la_4, \lb_1] \over \dotl[\la_4, \la_3] \dotl[\la_2, \lb_3]} + 1} \\ 
&= -{\dotl[\la_4, \la_3] \dotlm[\la_2, \lb_3]  \over \dotl[\la_2, \la_1] \dotlm[\la_4, \lb_1] + \dotl[\la_4, \la_3] \dotlm[\la_2, \lb_3]} \\
&= -{\dotl[\la_4, \la_3] \dotlm[\la_2, \lb_3] \over  i \dotl[\la_2, \la_4] E^{12,34}}.
\end{split}
\ee
Similarly, 
\be
{1 \over {w_{s_1}^+ \over w_{t_1}^+} - 1} = {\dotlm[\la_2, \lb_1] \dotl[\la_4, \la_1] \over \dotl[\la_2, \la_4] \dotlm[\la_1, \lb_1]},
\ee
and
\be
\label{pwfact3}
{1 \over {w_{s_1}^+ \over w_{t_2}^+} - 1} = -{\dotlm[\la_4, \lb_3] \dotl[\la_2, \la_3] \over i \dotl[\la_2, \la_4] E^{123,4}}.
\ee

We can now substitute \eqref{pwfact1} -- \eqref{pwfact3} in \eqref{fintermsofads} and \eqref{madsdef}
to recover our expression for ${\cal F}$.
\begin{align}
\label{firstexpsim} {\cal F} &= 
{E^{124,3} \over 4 \norm{k_3} \norm{k_4} \norm{k_1}} {\dotl[\la_2, \la_4] \dotlm[\la_4, \lb_3] \dotlm[\la_2, \lb_3] \dotlm[\la_2,\lb_1]  \over (E^{12, 34}) \dotl[\la_1, \la_2]} + \ldots
\end{align}
where the $\ldots$ indicate the various interchanges indicated in \eqref{secondexp} -- \eqref{tchannel}. This matches precisely with our
previous answer.

\subsection{Full MHV Amplitude}
To evaluate the full MHV amplitude we also need to consider
the (13)(24) partition, which we will call the ``u'' partition. With the value of $w$ being given by
\eqref{wlocation}, we can write down an expression for the integrand. 
\be
\label{tcurru}
\begin{split}
&{\cal I}_{u} = \sum_{\pm}  \Bigg\{
{R^{\text{YM}}(|\vect{k_1}|, |\vect{k_3}|, p) \over 2 \sqrt{2} |\vect{k_1}| |\vect{k_3}| p} {R^{\text{YM}}(|\vect{k_2}|, |\vect{k_4}|, p) \over 2 \sqrt{2}  |\vect{k_2}| |\vect{k_4}| p}  \\
&\begin{split} \times \Bigg[
&\bigl(\norm{k_1} + i p  - \norm{k_3} \bigr)^2 \bigl(i p + \norm{k_3} - \norm{k_1} \bigr) \bigl(\norm{k_1} + \norm{k_3} - i p \bigr) \bigl(\norm{k_4} + i p  - \norm{k_2} \bigr) \\ & \bigl(i p + \norm{k_2} - \norm{k_4} \bigr) \bigl(\norm{k_2} + \norm{k_4} + i p \bigr){\dotlb[\lb_{1}, \lb_{3}]^3 \over \dotlb[\lb_{1}, \lb_{\text{int}}] \dotlb[\lb_{3}, \lb_{\text{int}}] } {\dotl[\la_{4}, \la_{2}]^3 \over \dotl[\la_{2}, \la_{\text{int}}] \dotlb[\la_{4}, \la_{\text{int}}] } \\
&+ (E_p^{13}) (E_{-p}^{24}) \left(
\dotlb[\lb_1, \lb_3] \dotlb[\lb_1, \lb_{\text{int}}] \dotlb[\lb_3, \lb_{\text{int}}]\dotl[\la_2, \la_4] \dotlb[\la_2, \la_{\text{int}}] \dotlb[\la_4, \la_{\text{int}}] \right) 
\Bigg]  \Bigg\}
\end{split}\\
&\times {1 \over 4} {i p \over \left(p^2 + (\vect{k_1} + \vect{k_3})^2\right)} {w^{\mp} \over w^{\pm} - w^{\mp}}.
\end{split}
\ee
We can now simplify this, as above by using the identities
\begin{align}
\label{pi3int1}
& \la_1 \lb_1 + \la_3 \lb_3 + \la_{\text{int}} \lb_{\text{int}} = i E_p^{1 3} \sigma^3, \\
\label{pi3int2}
&\la_2 \lb_2 + \la_4 \lb_4 - \la_{\text{int}} \lb_{\text{int}}= i E_{-p}^{2 4} \sigma^3. 
\end{align}
Rewriting terms that involve $| \la_{\text{int}} \rangle \langle \lb_{\text{int}}| $ using
this, we find:
\begin{align}
\label{ipi3}
{\cal I}_{u} &=
 \sum_{\pm} \Bigg\{{-i \over 16 \pi \norm{k_1} \norm{k_2} \norm{k_3} \norm{k_4}} \Bigg[
{\bigl(\norm{k_1} + i p  - \norm{k_3} \bigr) \bigl(i p + \norm{k_3} - \norm{k_1} \bigr) \bigl(\norm{k_4} + i p  - \norm{k_2} \bigr)  \over (E_p^{13}) (E_{-p}^{24})}\\
\label{ipi3h+} 
&\times { \bigl(i p + \norm{k_2} - \norm{k_4} \bigr) \dotlb[\lb_{1}, \lb_{3}]^3 \dotl[\la_{2}, \la_{4}]^3 \over \left(\dotlb[\lb_{1}, \lb_{3}] \dotlb[\la_{2}, \la_{3}(w^{\pm}(p)] + i E_p^{13} \dotlm[\lb_1, \la_2] \right)  \left(\dotlb[\lb_{3}, \lb_{1}] \dotl[\la_{4}, \la_{1}(w^{\pm}(p))] + i E_p^{13} \dotlm[\lb_3, \la_4]\right)} \\
\label{ipi3h-1}
&+ {\dotlb[\lb_1, \lb_3]  \dotl[\la_2, \la_4] \left(\dotlb[\lb_3, \lb_{2}(w^{\pm})] \dotl[\la_4, \la_2] + i E_{-p}^{24} \dotlm[\lb_3, \la_4] \right) \over (\norm{k_1} + \norm{k_3} - i p) (\norm{k_2} + \norm{k_4} + i p) } \\ 
\label{ipi3h-2}
&\times \left(\dotlb[\lb_1, \lb_4(w^{\pm})] \dotl[\la_2, \la_4] + i E_{-p}^{24} \dotlm[\lb_1, \la_2] \right) \Bigg]  
\times {1 \over p^2 + (\vect{k_1} + \vect{k_3})^2} {w^{\mp} \over w^{\pm} - w^{\mp}}\Bigg\}. 
\end{align}

We now see that at the pole $E_p^{13} = 0$, only the term with $h_{\text{int}} = -1$ contributes with
\be
\label{tu1curr}
\begin{split}
 T_{u_1} &=  (2 \pi i) \underset{p = i (\norm{k_1} + \norm{k_3})}{\text{Res}} \Big[{\cal I}_{u} \Big] \\ = &{-i \over 2}
 {E^{123,4} E^{134,2} \over  E^T |\vect{k_{4}}|  |\vect{k_{2}}|} \sum_{\pm} { \dotl[\la_{2}, \la_{4}]^3 \over\dotlb[\la_{2}, \la_{3}(w_{u_1}^{\pm})]  \dotl[\la_{4}, \la_{1}(w_{u_1}^{\pm})]  \dotl[\l_1, \l_3] } {w_{u_1}^{\mp} \over w_{u_1}^{\pm} - w_{u_1}^{\pm}}.
\end{split}
\ee
Here $w_{u_1}^{\pm}$ are the two solutions to the quadratic equation
\be
\dotl[\la_1(w_{u_1}^{\pm}), \la_3(w_{u_1}^{\pm})] = 0.
\ee
Although these individual solutions involve square roots as we have pointed out above, we are always summing over both solutions, which gets rid of all the roots and leaves us with a rational function of the spinors. The reader may, if she prefers, easily use this to rewrite \eqref{tu1curr} as a rational function of the un-extended spinors.  

From the formula for the integrand above, it is easier to extract the residue at $p = -i(\norm{k_2} + \norm{k_4})$, which is, in any case, the same as the residue at $p = i (\norm{k_2} + \norm{k_4})$ since the integrand is even in $p$. To do this, we write \eqref{ipi3h+} using \eqref{pi3int2}. (We can also extract the residue at $E_p^{24}$ directly using \eqref{ipi3h-1} and \eqref{ipi3h-2} but that is less convenient.)
When we do this we find that 
\be
\begin{split}
T_{u_2} &= (2 \pi i) \underset{p = i (\norm{k_2} + \norm{k_4})}{\text{Res}} \Big[{\cal I}_{u} \Big] = \Big[T_{u_1} \Big]_{1 \leftrightarrow \bar{2}, 3 \leftrightarrow \bar{4}} \\ &= {-i \over 2}
 {E^{124,3} E^{234,1} \over  E^T |\vect{k_{3}}|  |\vect{k_{1}}|} \sum_{\pm} { \dotl[\lb_{1}, \lb_{3}]^3 \over\dotlb[\lb_{1}, \lb_{4}(w_{u_2}^{\pm})]  \dotl[\lb_{3}, \lb_{2}(w_{u_2}^{\pm})]  \dotl[\la_2, \la_4] } {w_{u_2}^{\mp} \over w_{u_2}^{\pm} - w_{u_2}^{\mp}}.
\end{split}
\ee
where $w_{u_2}^{\pm}$ are defined by the quadratic equation $\dotl[\lb_2(w_{u_2}^{\pm}), \lb_4(w_{u_4}^{\pm})] = 0$.

The full amplitude also involves the pole at $p = i \norm{k_1 + k_3}$. This is given by
\be
\begin{split}
&T_{u_3} = (2 \pi i) \underset{p = i \norm{k_1 + k_3}}{\text{Res}} \Big[{\cal I}_{u} \Big] = {i \over 16  \norm{k_1} \norm{k_2} \norm{k_3} \norm{k_4} \norm{k_1 + k_3}} \\ 
&\times \Bigg[
 { E^{3s,1} E^{1s,3} E^{4,s2} E^{2,s4} \dotlb[\lb_{1}, \lb_{3}]^3 \dotl[\la_{2}, \la_{4}]^3 \over E^{13,s} E^{24s} \left(\dotlb[\lb_{1}, \lb_{3}] \dotlb[\la_{2}, \la_{3}] + i E^{13,s} \dotlm[\lb_1, \la_2] \right)  \left(\dotlb[\lb_{3}, \lb_{1}] \dotl[\la_{4}, \la_{1}] + i E^{13,s} \dotlm[\lb_3, \la_4]\right)} \\
&+ {\dotlb[\lb_1, \lb_3]  \dotl[\la_2, \la_4] \left(\dotlb[\lb_3, \lb_{2}] \dotl[\la_4, \la_2] + i E^{2 4 s}  \dotlm[\lb_3, \la_4] \right)  \left(\dotlb[\lb_1, \lb_4] \dotl[\la_2, \la_4] + i E^{24s} \dotlm[\lb_1, \la_2] \right) \over E^{13s} E^{24,s} }  \Bigg].  
\end{split}
\ee

The full contribution of this partition is given by $T_{u_1} + T_{u_2} + T_{u_3}$. 

To get the full MHV current correlator we just need to add the contributions from the $s$ and $t$ partitions that we have already computed above. One note of caution is that each of these terms now comes with the appropriate color-factor. For example, the $s$-channel partition is multiplied by the color-factor $f^{12e} f^{e34}$ and similarly for the $t$ and $u$ channels. 

We see now that the flat space limit is manifest. In analogy to the color ordered correlators, let us define:
\be
\label{fullmhvcurr}
\begin{split}
{\cal M}^{\text{AdS}}(w) = &\left(M^{\text{MHV}}(w) P^+(w)  + M^{\overline{\text{MHV}}}(w) P^{-}(w)\right) f^{1 2 e}f^{e 3 4} \\ &+ \left(\tilde{M}^{\text{MHV}}(w) \tilde{P}^+(w) + \tilde{M}^{\overline{\text{MHV}}} \tilde{P}^{-}(w) \right) f^{1 4 e} f^{e 2 3}.
\end{split}
\ee
Here
\begin{equation}
\label{mhvbarmhvfull}
\begin{split}
&\tilde{M}^{\text{MHV}}(w) = {\dotl[\la_2, \la_4]^3 \over \dotl[\la_1(w), \la_4] \dotl[\la_2, \la_3(w)] \dotl[\la_3(w), \la_1(w)]}, \\ 
 &\tilde{M}^{\overline{\text{MHV}}}(w) = {\dotl[\lb_1, \lb_3]^3 \over \dotl[\lb_1, \lb_4(w)] \dotl[\lb_4(w), \lb_2(w)] \dotl[\lb_2(w), \lb_3]},
\end{split}
\end{equation}
If we adopt the notation:
\be
w_{y_1}^+ = w_{u_1}^+; \quad w_{y_2}^+ = w_{u_1}^-; w_{y_1}^{-} = w_{u_2}^+; w_{y_2}^{-} = w_{u_2}^{-},
\ee
then we can define the $\tilde{P}$ functions using 
\be
\begin{split}
\tilde{P}^{+}(w) &\equiv  
{(w - w_{y_2}^{+}) (w - w_{t_1}^{+}) (w - w_{t_2}^{+}) \over (w_{y_1}^{+} - w_{y_2}^{+}) (w_{y_1}^{+} - w_{t_1}^{+}) (w_{y_1}^{+} - w_{t_2}^{+})} {E^{123,4} E^{134,2} \over 2 \norm{k_2} \norm{k_4}}
\\ &+  {(w - w_{y_1}^{+}) (w - w_{t_1}^{+}) (w - w_{t_2}^{+}) \over (w_{y_2}^{+} - w_{y_1}^{+}) (w_{y_2}^{+} - w_{t_1}^{+}) (w_{y_2}^{+} - w_{t_2}^{+})} {E^{123,4} E^{134,2} \over 2 \norm{k_2} \norm{k_4}} \\ &+ 
{(w - w_{y_1}^{+}) (w - w_{y_2}^{+}) (w - w_{t_1}^{+}) \over (w_{t_1}^{+} - w_{y_2}^{+}) (w_{t_1}^{+} - w_{y_2}^{+}) (w_{t_1}^{+} - w_{t_2}^{+})} {E^{124,3} E^{134,2} \over 2 \norm{k_3} \norm{k_2}} \\ &+  
{(w - w_{y_1}^{+}) (w - w_{y_2}^{+}) (w - w_{t_1}^{+}) \over (w_{t_2}^{+} - w_{y_1}^{+}) (w_{t_2}^{+} - w_{y_2}^{+}) (w_{t_2}^{+} - w_{t_1}^{+})} {E^{234,1} E^{123,4} \over 2 \norm{k_1} \norm{k_4}}.
\end{split}
\ee
The other interpolating function $\tilde{P}^{-}$ is defined similarly:
\be
\begin{split}
\tilde{P}^{-}(w) &\equiv  
{(w - w_{y_2}^{-}) (w - w_{t_1}^{-}) (w - w_{t_2}^{-}) \over (w_{y_1}^{-} - w_{y_2}^{-}) (w_{y_1}^{-} - w_{t_1}^{-}) (w_{y_1}^{-} - w_{t_2}^{-})} {E^{124,3} E^{234,1} \over 2 \norm{k_1} \norm{k_3}}
\\ &+  {(w - w_{y_1}^{-}) (w - w_{t_1}^{-}) (w - w_{t_2}^{-}) \over (w_{y_2}^{-} - w_{y_1}^{-}) (w_{y_2}^{-} - w_{t_1}^{-}) (w_{y_2}^{-} - w_{t_2}^{-})} {E^{124,3} E^{234,1} \over 2 \norm{k_1} \norm{k_3}} \\ &+ 
{(w - w_{y_1}^{-}) (w - w_{y_2}^{-}) (w - w_{t_1}^{-}) \over (w_{t_1}^{-} - w_{y_2}^{-}) (w_{t_1}^{-} - w_{y_2}^{-}) (w_{t_1}^{-} - w_{t_2}^{-})} {E^{124,3} E^{134,2} \over 2 \norm{k_3} \norm{k_2}} \\ &+  
{(w - w_{y_1}^{-}) (w - w_{y_2}^{-}) (w - w_{t_1}^{-}) \over (w_{t_2}^{-} - w_{y_1}^{-}) (w_{t_2}^{-} - w_{y_2}^{-}) (w_{t_2}^{-} - w_{t_1}^{-})} {E^{234,1} E^{123,4} \over 2 \norm{k_1} \norm{k_4}}.
\end{split}
\ee

It is this term evaluated at $w = 0$ --- ${\cal M}^{\text{AdS}}(0)$  --- that plays the role that ${\cal F}$ played
in \eqref{colororderedmhv}. 
Just as in the case above, near $E^T = 0$, we find that $\tilde{P}^{\pm} = 2$.
So the full MHV amplitude (including contributions from other partitions) goes like:
\[
{4 \over E^T} \Big({ M^{\text{MHV}}(0) f^{1 2 e}f^{e 3 4}  +  \tilde{M}^{\text{MHV}}(0)  f^{1 4 e} f^{e 2 3}}\Big) + \ldots
\]
where $\ldots$ are terms that are non-singular at $E^T = 0$. This is exactly what we need.

\section{MHV Correlators for the Stress Tensor \label{secmhvgrav}}
We now turn to correlation functions of the stress tensor. Our objective
in this section is to write out the formula \eqref{formulafinal} explicitly in terms of spinors and check that the MHV graviton amplitude
appears in the flat space limit. We will achieve this in two steps.
First we expand out the integrand that appears in \eqref{formulafinal}. Then we expand out the residues that appear in that formula. The procedure for writing down the integrand is almost identical to the case of 
conserved currents. However, there is one important difference in the final result; this is the fact that the poles that appear from the three point amplitudes are now double poles. Consequently, to extract the residue we need to take a derivative. This complicates our final formulas.

We will carry out this procedure with the same configuration of external
polarizations that we used for conserved currents. Namely, we will take
$\{h_1, h_2, h_3, h_4\} = \{1,-1,1,-1\}$. The generalization to 
other helicity configurations involves a straightforward procedure
that is very similar to the one that we present in detail below. The
reader may also use the computer program that accompanies this paper
to generate answers for any combination of external helicities either 
analytically or numerically. 

We will first write down the integrands for the three different partitions that appear in \eqref{formulafinal}. These are the (12)(34) partition,
the (14)(23) partition, and the (13)(24) partition. In each case, we then 
extract the residue at the poles specified in \eqref{polelist}

\subsection{(12)(34) Partition: \\}
Let us start by considering the (12)(34) partition. 
We can write down an expression for the $p$-integrand corresponding
to this partition, using the 
three-point amplitudes above. This expression is given by
\be
\label{tpi1prelim}
\begin{split}
{\cal I}_{s} = \sum_{\pm} &\Bigg\{{R^{\text{gr}}(|\vect{k_1}|, |\vect{k_2}|, p) \over 32  |\vect{k_1}|^2 |\vect{k_2}|^2 p^2} \bigl(\norm{k_2} + i p  - \norm{k_1} \bigr)^2 \bigl(i p + \norm{k_1} - \norm{k_2} \bigr)^2 \bigl(\norm{k_1} + \norm{k_2} - i p \bigr)^2 \\
&{R^{\text{gr}}(|\vect{k_3}|, |\vect{k_4}|, -p) \over 32  |\vect{k_3}|^2 |\vect{k_4}|^2 p^2} \bigl(\norm{k_4} - i p  - \norm{k_3} \bigr)^2 \bigl(-i p + \norm{k_3} - \norm{k_4} \bigr)^2 \bigl(\norm{k_3} + \norm{k_4} + i p \bigr)^2 \\
&\begin{split} \times \Bigg[ &\left({\dotlb[\lb_{1}, \lb_{\text{int}}]^3 \over \dotlb[\lb_{1}, \lb_{2}(w_{s}^{\pm})] \dotlb[\lb_{2}(w_{s}^{\pm}), \lb_{\text{int}}] } {\dotl[\la_{4}, \la_{\text{int}}]^3 \over \dotlb[\la_{3}(w_{s}^{\pm}), \la_{4}] \dotlb[\la_{3}(w_{s}^{\pm}), \la_{\text{int}}] }\right)^2 \\ &+ \left({\dotl[\la_2, \la_{\text{int}}]^3 \over \dotl[\la_{1}(w_{s}^{\pm}), \la_{2}] \dotl[\la_{1}(w_{s}^{\pm}), \la_{\text{int}}] } {\dotlb[\lb_{3}, \lb_{\text{int}}]^3 \over \dotlb[\lb_{3}, \lb_{4}(w_{s}^{\pm})] \dotlb[\lb_{4}(w_{s}^{\pm}), \lb_{\text{int}}] }\right)^2
\Bigg] 
\end{split}\\
&\times {i p  \over 8\left( p^2 + (\vect{k_1} + \vect{k_2})^2\right)} {w_{s}^{\mp} \over w_{s}^{\pm} - w_{s}^{\mp}}  \Bigg\}.
\end{split}
\ee
Note that if, on the left hand side, we use the norm $\norm{k_{\text{int}}} = i p$, on the right hand side we need to use $(-i p)$. We also
need to flip the sign of $\la_{\text{int}}$, but this does not matter
because of the fact that the integrand involves only the ``square''
of this term. By expanding out $R^{\text{gr}}$ and recalling the identity \eqref{isintspin}, we find 
\be
\label{tpi1}
\begin{split}
&{\cal I}_{s} = \sum_{\pm}  {1 \over 2^{12} \pi} \Bigg\{\frac{\left(|\vect{k_{1}}|^2+4 \norm{k_{2}} \norm{k_{1}} +|\vect{k_{2}}|^2+p^2\right) \left(|\vect{k_{3}}|^2+4 \norm{k_{4}} \norm{k_{3}} +|\vect{k_{4}}|^2+p^2\right) }{(|\vect{k_{1}}|  |\vect{k_{2}}| |\vect{k_{3}}|  |\vect{k_{4}}| )^{2} \left(|\vect{k_{1}}|+|\vect{k_{2}}|+ i p \right)^2  \left(|\vect{k_{3}}|+|\vect{k_{4}}|- i p \right)^2} \\ &\times   \bigl(\norm{k_2} + i p  - \norm{k_1} \bigr)^2 \bigl(i p + \norm{k_1} - \norm{k_2} \bigr)^2  \bigl(\norm{k_3} + i p  - \norm{k_4} \bigr)^2 \bigl(i p + \norm{k_4} - \norm{k_3} \bigr)^2\\
&\begin{split} 
&\times \Bigg[{ 
\Big(
\dotlb[\lb_1, \lb_2(w_{s}^{\pm})] 
\dotl[\la_{4}, \la_2] + i E_p^{12} 
\dotlm[\lb_1,\la_4] \Big)^6
  \over \dotlb[\lb_{1}, \lb_{2}(w_{s}^{\pm})]^2 
\left(\dotlb[\lb_{2}(w_{s}^{\pm}), \lb_{1}] \dotl[\la_1(w_{s}^{\pm}), \la_3(w_{s}^{\pm})] - i E_p^{12} \dotlm[\la_3(w_{s}^{\pm}), \lb_2(w_{s}^{\pm})] \right)^2 \dotl[\la_{4}, \la_{3}(w_{s}^{\pm})]^2 }
\\ &
+{\left(\dotl[\la_2, \la_1(w_{s}^{\pm})]\dotlb[\lb_{3}, \lb_{1}] 
+ i E_p^{12} \dotlm[\la_2, \lb_3] \right)^6  \over 
\dotl[\la_{1}(w_{s}^{\pm}), \la_{2}]^2 \left(\dotl[\la_{1}(w_{s}^{\pm}), \la_2] \dotlb[\lb_{4}(w_{s}^{\pm}), \lb_{2}(w_{s}^{\pm})] + i E_p^{12} \dotlm[\la_1(w_{s}^{\pm}),\lb_4(w_{s}^{\pm})] \right)^2 \dotlb[\lb_{3}, \lb_{4}(w_{s}^{\pm})]^2 }
\Bigg] 
\end{split}
\\ &\times { i \over p^2 + (\vect{k_1} + \vect{k_2})^2} {w_{s}^{\mp} \over w_{s}^{\pm} - w_{s}^{\mp}}  \Bigg\},
\end{split}
\ee
where we have remind the reader that $E_p^{n m} \equiv i p + \norm{k_n} + \norm{k_m}.$

\paragraph{Extracting the Residues: }
We now proceed to implement \eqref{formulafinal} and extract the 
residue from the integrand above. An important difference from the conserved current computation is that, as we mentioned above, the poles that appear from three point amplitudes are double poles. So extracting the residue involves taking a derivative of the integrand with respect to $p$ at the pole. 

It seems more convenient to perform this procedure through ``logarithmic differentiation'': first we write down an expression for the value of the integrand, with the singular term stripped off, at the pole. Next we write down an expression for the {\em ratio} of the derivative of this term to the term itself. 

The procedure of extracting the value of the integrand is almost identical to the one we followed to obtain \eqref{firstexp1}. The difference is that various terms are squared. When $E_{p}^{12} = 0$ the value of $w^{\pm}(p)$ for this partition is still defined by
\eqref{ws1p} and \eqref{ws1m}.  To simplify the integrand we throw away all terms proportional to $E_p^{12}$. Apart from \eqref{explicitspinsws1p},
we need one more explicit expression for an extended spinor:
\be
\label{explicitspinsws1p2}
 \lb_2(w_{s1}^+) =  \lb_3 {\dotlm[\la_2,\lb_2] + \dotlm[\la_1,\lb_1] - i E^T\over \dotlm[\la_2,\lb_3]} - {\dotlm[\la_1,\lb_1] \over \dotlm[\la_2,\lb_1]} \lb_1 + {i E^T  \over \dotl[\la_2, \la_4]} \lad_4.
\ee

From these expressions, we find
 \be
 \label{iplus1}
 \begin{split}
 {\cal V}^{+}_{s_1} =  &(2 \pi i) \lim_{E_p^{12} \rightarrow 0} \left(E_p^{12}\right)^2 {\cal I}_s  \\ = &\Big[\frac{-i \norm{k_{2}}  \left(E^{34,12} (E^T) 
  + 2 \norm{k_{3}} \norm{k_{4}} \right)  (E^{3,124})^2  }{256 \norm{k_{1}}  \norm{k_{3}}^2  \norm{k_{4}}^{2}  (E^T)^2  (E^{34,12})^2 \dotl[\la_1, \la_2]}\Big]  \dotl[\la_2, \la_4] \dotlm[\la_2, \lb_1]^2   \\ 
&\times  \left(\dotlb[\lb_1, \lb_3]  \dotl[\la_2, \la_4] E^{12,34} +  E^T \dotlm[\lb_1, \la_4] \dotlm[\la_2, \lb_3] \right) \dotlm[\la_4, \lb_3]^2 \dotlm[\la_2, \lb_3].
\end{split}
\ee

We now turn to an evaluation of the derivative. As we mentioned above
it is convenient to work with the quantity:
\be
{\cal D}_{s_1}^{+} = \lim_{E_p^{12} \rightarrow 0} {d \over d p} \log \big[ \left(E_p^{12}\right)^2 {\cal I}_s \big], 
\ee
where it is understood that the limit is taken at the value of $w^{\pm}(p)$ in \eqref{tpi1} corresponding to $w_{s_1}^+$. 

Before we evaluate this expression, note that we can also define $w^{\pm}$
through
\be
 \dotl[\la_1(w_{s}), \la_2] \dotl[\lb_1, \lb_2(w_{s})] = -(\norm{k_1} + \norm{k_2} - i p) E_p^{12}.
\ee
At a pole in the $p$-integral, where $E_p^{12} = 0$, one of the terms on the left hand side must vanish. We have defined $w_{s_1}^{+}$ to 
be the pole where the first dot product vanishes and $w_{s_1}^{-}$ to 
be the pole where the second dot product vanishes. With a slight abuse of notation, defining $ {d w^{\pm} \over d p} 
\equiv {d w_{s_1}^{\pm} \over d p}$ at this point,  we have
\begin{align}
&\beta_1 {d w^+_{s_1} \over d p} \dotlm[\lb_1, \la_2]  \dotl[\lb_1, \lb_2(w_{s}^+)] = {-2 i(\norm{k_1} + \norm{k_2})}  \\
\label{gamma1pdef}
&\Rightarrow \gamma_1^{+} \equiv {d w^+_{s_1} \over d p}  = {-2 i (\norm{k_1} + \norm{k_2}) \over \beta_1  \dotlm[\lb_1, \la_2]  \dotl[\lb_1, \lb_2(w_{s_1}^+)]}  = {- 2 i (\norm{k_1} + \norm{k_2}) \over  \dotlm[\lb_1, \la_2] \left(\beta_1 \dotl[\lb_1,\lb_2] + \beta_2 \dotl[\la_1, \la_2]\right)}.
\end{align}
Furthermore
\begin{align}
&\beta_2 {d w^-_{s_1} \over d p}   \dotl[\la_1(w_{s}), \la_2] \dotlm[\lb_1, \la_2] = {-2 i(\norm{k_1} + \norm{k_2})}  \\
\label{gamma1mdef}
&\Rightarrow \gamma_1^{-}  \equiv {d w^-_{s_1} \over d p}  =  {-2 i(\norm{k_1} + \norm{k_2}) \over \beta_2  \dotl[\la_1(w_{s_1}^-), \la_2] \dotlm[\lb_1, \la_2]} = -\gamma_1^+.
\end{align}
With this observation and notation,  ${\cal D}_{s_1}^{+}$ is given by
\be
\label{dsp}
\begin{split}
&{\cal D}_{s_1}^{+} = 2 i \left( \norm{k_1} + \norm{k_2}\right)\left(
{1 \over E^{34,12} E^T + 2 \norm{k_3} \norm{k_4}} + {1 \over \dotl[\la_1, \la_2] \dotl[\lb_1, \lb_2]} \right) \\ &+  i \left(
{2 \over E^{3,124}} + {2 \over E^{4,123}} + {2 \over E^T} \right) -{1 \over \dotlb[\lb_1, \lb_2(w_{s_1}^+)] } \left({6 \dotlm[\lb_1, \la_4] \over \dotl[\la_4, \la_2]}   - {2 \dotlm[\la_3(w_{s_1}^+), \lb_2(w_{s_1}^+)]  \over \dotl[\la_1(w_{s_1}^+), \la_3(w_{s_1}^+)]} \right) \\ 
&+ 2 \gamma_1^+ \Bigg\{ {\beta_2 \dotlm[\lb_1, \la_2]  \over \dotlb[\lb_1, \lb_2(w_{s_1}^+)]} 
-   {\beta_3 \dotlm[\lb_3, \la_4]  \over \dotl[\la_4,\la_3(w_{s_1}^+)]}
+  {\beta_1 \dotlm[\lb_1, \la_3(w_{s_1}^+)] - \beta_3 \dotlm[\la_1(w_{s_1}^+), \lb_3] \over \dotl[\la_1(w_{s_1}^+), \la_3(w_{s_1}^+)]} \Bigg\} \\
 &- \gamma_1^{+} {w_{s_1}^+ + w_{s_1}^- \over w_{s_1}^{-} \left(w_{s_1}^+ - w_{s_1}^{-}\right)}  .
\end{split}
\ee
If the reader wishes to expand this expression out in terms of un-extended spinors, she can do so using \eqref{explicitspinsws1p}, \eqref{explicitspinsws1p2}
and the additional identity:
\be
\la_3(w_{s1}^+) = {\dotlm[\la_3,\lb_3] + \dotlm[\la_4,\lb_4] - i E^T\over \dotlm[\la_2,\lb_3]} \la_2 - \la_4 {\dotlm[\la_4,\lb_4] - i E^T \over \dotlm[\la_4,\lb_3]}.
\ee
However, this does not provide much additional insight so we have left 
the expression above as is.

Note that the contribution of ${\cal V}^{-}_{s_1} {\cal D}^{-}_{s_1}$, which is just the residue at $E_p^{12} = 0$ but with $w^{\pm}(p) = w_{s_1}^{-}$
in \eqref{tpi1} can be easily incorporated, just through some substitutions. We have
\be
\label{t11result}
T_{s_1} = (2 \pi i) \underset{p = i (\norm{k_1} + \norm{k_2})}{\text{Res}}\big[{\cal I}_s \big] =  {\cal V}_{s_1}^{+} {\cal D}_{s_1}^{+} + (\la_1, \la_2, \la_3, \la_4) \leftrightarrow (\lb_2, \lb_1, \lb_4, \lb_3).
\ee

There is another pole in ${\cal I}_s$ at $E_p^{34} = 0$.\footnote{Although the
pole that is manifest in \eqref{tpi1} is actually at $E_{-p}^{34} = 0$, we should remember that the integrand is even in $p$ and we can rewrite
it to make the pole at $E_{p}^{34} = 0$ manifest instead.} The contribution of this pole is just obtained by interchanging $(12)$ with $(34)$:
\be
\label{t12result}
T_{s_2} = \Big[T_{s_1}\Big]_{1 \leftrightarrow 3, 2 \leftrightarrow 4}.
\ee

We now turn to the second kind of contribution to the correlator from this partition,  which comes from 
the pole where the denominator of the propagator vanishes. Fortunately,
this is a simple pole! This pole occurs at:
\be
p =  i \norm{k_1 + k_2}.
\ee
Note that the factor ${w^{+} \over w^{+} - w^{-}}$ becomes the identity,
since $w^{-} = 0$. Also, we do not need the other pole $w^{+}$ at all, 
since that contribution vanishes because of this factor (which, in that
case, would be ${w^{-} \over w^{+} - w^{-}}$). We have already seen this 
in the calculation for current-correlators.
Picking up the pole at $w = 0$, we have
\be
\label{t13result}
\begin{split}
&T_{s_3} =  (2 \pi i) \underset{p = i \norm{k_1 + k_2}}{\text{Res}} {\cal I}_s \\ &=\frac{-i \left(E^{12,s}E^{12s} + 2 \norm{k_{2}} \norm{k_{1}} \right) \left(E^{34,s}E^{34s}+2\norm{k_{4}} \norm{k_{3}}\right) (E^{1s,2})^2 (E^{2s,1})^2 (E^{3s,4})^2 (E^{4s,3})^2 }{  2^{12} (|\vect{k_{1}}|  |\vect{k_{2}}| |\vect{k_{3}}|  |\vect{k_{4}}| )^{2} (E^{12,s})^2  (E^{34s})^2 \norm{k_1 + k_2}} \\
&\begin{split} 
\times \Bigg[&{ 
\Big(
\dotlb[\lb_1, \lb_2] 
\dotl[\la_{4}, \la_2] + i E^{12,s} 
\dotlm[\lb_1,\la_4] \Big)^6
  \over \dotlb[\lb_{1}, \lb_{2}]^2 
\left(\dotlb[\lb_{2}, \lb_{1}] \dotl[\la_1, \la_3] - i E^{12,s} \dotlm[\la_3, \lb_2] \right)^2 \dotlb[\la_{3}, \la_{4}]^2 }
\\ &
+{\left(\dotl[\la_2, \la_1]\dotlb[\lb_{3}, \lb_{1}] 
+ i E^{12,s} \dotlm[\la_2, \lb_3] \right)^6  \over 
\dotl[\la_{1}, \la_{2}]^2 \left(\dotl[\la_{1}, \la_2] \dotlb[\lb_{4}, \lb_{2}] + i E^{12,s} \dotlm[\la_1, \lb_4] \right)^2 \dotlb[\lb_{3}, \lb_{4}]^2 }
\Bigg], 
\end{split}
\end{split}
\ee
where we remind the reader that $s$ stands for sum and so, for example,
\be
E^{12,s} \equiv \norm{k_1} + \norm{k_2} - \norm{k_1 + k_2}.
\ee

To sum up the contribution of the $(12)(34)$ partition is given by
\be
\label{ts}
T_s = T_{s_1} + T_{s_2} + T_{s_3},
\ee
where the three terms on the right are given by \eqref{t11result}, \eqref{t12result} and  \eqref{t13result}.

\subsection{(14)(23) Partition \\}
This leads to the same expression as above with the substitution
$2 \leftrightarrow 4$. The computation of the residue is exactly the same as the one for the (12)(34) partition with $4 \leftrightarrow 2$:
\be
\label{tt}
T_t = \big[ T_s \big]_{\la_4 \leftrightarrow \la_2, \lb_4 \leftrightarrow \lb_2}.
\ee

\subsection{(13)(24) Partition \\}
This partition has a slightly different structure. We can write
the integrand for this partition as ${\cal I}_u = \sum_{\pm} {\cal I}^{\pm}_u$, corresponding to the two different values of $w$ where 
\be
\label{tpi2}
\begin{split}
&{\cal I}_{u}^{\pm} = 
{R^{\text{gr}}(|\vect{k_1}|, |\vect{k_3}|, p) \over 32 |\vect{k_1}|^2 |\vect{k_3}|^2 p^2} {R^{\text{gr}}(|\vect{k_2}|, |\vect{k_4}|, p) \over 32  |\vect{k_2}|^2 |\vect{k_4}|^2 p^2} \times { i p \over 8 \left(p^2 + (\vect{k_1} + \vect{k_3})^2\right)} {w_{u}^{\pm} \over w_{u}^{\pm} - w_{u}^{\mp}}. \\
&\begin{split} \times \Bigg[
&\bigl(\norm{k_1} + i p  - \norm{k_3} \bigr)^2 \bigl(i p + \norm{k_3} - \norm{k_1} \bigr)^2 \bigl(\norm{k_1} + \norm{k_3} - i p \bigr)^2 \bigl(\norm{k_4} + i p  - \norm{k_2} \bigr)^2 \\ & \bigl(i p + \norm{k_2} - \norm{k_4} \bigr)^2 \bigl(\norm{k_2} + \norm{k_4} + i p \bigr)^2 \left({\dotlb[\lb_{1}, \lb_{3}]^3 \over \dotlb[\lb_{1}, \lb_{\text{int}}] \dotlb[\lb_{3}, \lb_{\text{int}}] } {\dotl[\la_{4}, \la_{2}]^3 \over \dotl[\la_{2}, \la_{\text{int}}] \dotlb[\la_{4}, \la_{\text{int}}] }\right)^2 \\
&+ (E_p^{13})^2 (E_{-p}^{24})^2 \left(
\dotlb[\lb_1, \lb_3] \dotlb[\lb_1, \lb_{\text{int}}] \dotlb[\lb_3, \lb_{\text{int}}]\dotl[\la_2, \la_4] \dotlb[\la_2, \la_{\text{int}}] \dotlb[\la_4, \la_{\text{int}}] \right)^2 
\Bigg]. 
\end{split}\\
\end{split}
\ee
We can write this expression as
\be
\label{tpi3}
\begin{split}
&{\cal I}_{u}^{\pm} =  \frac{\left(|\vect{k_{1}}|^2+4 \norm{k_{3}} \norm{k_{1}} +|\vect{k_{3}}|^2+p^2\right) \left(|\vect{k_{2}}|^2+4 \norm{k_{2}} \norm{k_{4}} +|\vect{k_{4}}|^2+p^2\right) }{ 2^{12} \pi (|\vect{k_{1}}|  |\vect{k_{4}}| |\vect{k_{3}}|  |\vect{k_{2}}| )^{2}}
\\
&\begin{split} \times \Bigg[
&{\bigl(\norm{k_1} + i p  - \norm{k_3} \bigr)^2 \bigl(i p + \norm{k_3} - \norm{k_1} \bigr)^2 \bigl(\norm{k_4} + i p  - \norm{k_2} \bigr)^2 \bigl(i p + \norm{k_2} - \norm{k_4} \bigr)^2 \over (E_p^{13})^2 (E_{-p}^{24})^2 }\\
&\times {\dotlb[\lb_{1}, \lb_{3}]^6 \dotl[\la_{4}, \la_{2}]^6 \over \left(\dotlb[\lb_{1}, \lb_{3}] \dotlb[\la_{2}, \la_{3}(w_u^{\pm})] + i E_p^{13} \dotlm[\lb_1, \la_2] \right)^2  \left(\dotlb[\lb_{3}, \lb_{1}] \dotl[\la_{4}, \la_{1}(w_u^{\pm})] + i E_p^{13} \dotlm[\lb_3, \la_4]\right)^2} \\ 
&+ {\dotlb[\lb_1, \lb_3]^2  \dotl[\la_2, \la_4]^2 \over (\norm{k_1} + \norm{k_3} - i p)^2 (\norm{k_2} + \norm{k_4} + i p)^2 } \\ 
&\times  \left(\dotlb[\lb_3, \lb_{2}(w_{u}^{\pm}] \dotl[\la_4, \la_2] + i E_{-p}^{24} \dotlm[\lb_3, \la_4] \right)^2 \left(\dotlb[\lb_1, \lb_4(w_{u}^{\pm})] \dotl[\la_2, \la_4] + i E_{-p}^{24} \dotlm[\lb_1, \la_2] \right)^2 \Bigg]  
\\
\end{split}\\
&\times {i \over p^2 + (\vect{k_1} + \vect{k_3})^2} {w_{u}^{\mp} \over w_{u}^{\pm} - w_{u}^{\mp}}.
\end{split}
\ee

\paragraph{Extracting the Residues: }
Let us start by picking up the residue at $p = i(\norm{k_1} + \norm{k_3})$. We see from the expression above that only the term with $h_{\text{int}} = -1$ contributes to this residue; however, both values of $w$ are important. We have
\be
\begin{split}
&{\cal V}^{\pm}_{u_1} = (2 \pi i) \lim_{E_p^{13} \rightarrow 0} \left(E_p^{12}\right)^2 {\cal I}_u^{\pm} \\
&=  \frac{\norm{k_1} \norm{k_3}  \left(E^{24,13} E^T + 2 \norm{k_2} \norm{k_4} \right)  (E^{4,123})^2 (E^{2,134})^2 \dotlb[\lb_{1}, \lb_{3}] \dotl[\la_{4}, \la_{2}]^6}{ 64  (|\vect{k_{4}}| |\vect{k_{2}}| )^{2} (E^T)^2  \dotlb[\la_{2}, \la_{3}(w_u^{\pm})]^2 \dotl[\la_{4}, \la_{1}(w_u^{\pm})]^2 \dotl[\la_1, \la_3] }
{w_{u_1}^{\mp} \over w_{u_1}^{\pm} - w^{\mp}_{u_1}}.
\end{split}
\ee

Now we need the derivative of the log of the integrand. Recall that 
$w$ can be defined through
\be
\label{w3def}
\dotl[\la_1+ \lbd_1 \beta_1 w^{\pm}, \la_3 + \lbd_3 \beta_3 w^{\pm}] \dotl[\lb_1, \lb_3] = -\left((\norm{k_1} + \norm{k_3} - i p\right) E_p^{13},
\ee
and this also leads to an expression for the derivative when $E_p^{13} = 0$:
\be
\label{gamma3def}
\gamma_{3}^{\pm} \equiv {d w_{u_1}^{\pm} \over dp} = {2 i (\norm{k_1} + \norm{k_3}) \over   \dotl[\lb_1, \lb_3] \left(\beta_3 \dotlm[\la_1(w^\pm), \lb_3] - \beta_1 \dotlm[\la_3(w^{\pm}), \lb_1]\right) }.
\ee
We have $\gamma_3^+ = - \gamma_3^{-}$.

We can now evaluate the derivative that we need
\be
\label{du1}
\begin{split}
{\cal D}_{u_1}^{+} &= \lim_{E_p^{13} \rightarrow 0} {d \over d p} \log \big[ \left(E_p^{13}\right)^2 {\cal I}_u^+ \big] \\ 
&=  {2 i (\norm{k_1} + \norm{k_3}) \over E^{24,13} E^T + 2 \norm{k_2} \norm{k_4}} + {2 i (\norm{k_1} + \norm{k_3}) \over \dotl[\la_1, \la_3] \dotl[\lb_1, \lb_3]}  + {2 i  \over E^{4,123}} + {2 i \over E^{2,134}} + {2 i \over E^T} \\
& - \gamma_3^{+}\left({2 \beta_3 \dotlm[\la_2, \lb_3] \over \dotlb[\la_{2}, \la_{3}(w_u^{\pm})]} + {2 \beta_1 \dotlm[\la_4, \lb_1] \over  \dotl[\la_{4}, \la_{1}(w_u^{\pm})]} + {w_{u_1}^{+} + w_{u_1}^{-} \over w_{u_1}^{-}\left(w_{u_1}^+ - w_{u_1}^-\right)} \right) \\ &+  {2 \over  \dotlb[\lb_{1}, \lb_{3}]} \left({\dotlm[\lb_1, \la_2] \over \dotlb[\la_{2}, \la_{3}(w_u^{\pm})]} - {\dotlm[\lb_3, \la_4] \over \dotl[\la_{4}, \la_{1}(w_u^{\pm})]} \right). 
\end{split}
\ee

We can now write
\be
T_{u_1} = \sum_{\pm} {\cal V}^{\pm}_{u_1} {\cal D}^{\pm}_{u_1}. 
\ee
The residue at $p = i (\norm{k_2} + \norm{k_4})$ can be obtained by 
interchanges in the expression above:
\be
T_{u_2} = \Big[T_{u_1}\Big]_{1 \leftrightarrow \bar{2}, 3 \leftrightarrow \bar{4}}.
\ee
(It is understood that alongside we also take $\bar{1} \leftrightarrow 2, \bar{3} \leftrightarrow 4$.)

Finally, we turn to the contribution from the pole at $p^2 = -(\vect{k_2} + \vect{k_4})^2 = -(\vect{k_1} + \vect{k_3})^2$, which occurs at $w = 0$. This is a first order pole and we can evaluate it as above.
\be
\begin{split}
&T_{u_3} = (2 \pi i) \underset{p = i \norm{k_1 + k_3}}{\text{Res}} {\cal I}_u \\ 
&= \frac{-i \left(E^{13,s} E^{13s} + 2 \norm{k_1} \norm{k_3}\right) \left(E^{24,s}E^{24s} + 2 \norm{k_2} \norm{k_4} \right)}{ 2^{12} (|\vect{k_{1}}|  |\vect{k_{4}}| |\vect{k_{3}}|  |\vect{k_{2}}| )^{2} \norm{k_1 + k_3}  }\\
&\times \Bigg[
 {\bigl(E^{1s,3}\bigr)^2 \bigl(E^{3s,1}\bigr)^2 \bigl(E^{2s,4}\bigr)^2    \bigl(E^{24s}\bigr)^2 \dotlb[\lb_{1}, \lb_{3}]^2 \dotl[\la_{4}, \la_{2}]^6 \over \bigl(E^{13,s}\bigr)^2  \bigl(E^{24,s}\bigr)^2 \left(  \dotlb[\la_{2}, \la_{3}] + i E^{13,s} {\dotlm[\lb_1, \la_2] \over \dotl[\lb_1, \lb_3]}\right)^2  \left( \dotl[\la_{4}, \la_{1}] + i E^{13,s} {\dotlm[\lb_3, \la_4] \over \dotl[\lb_3, \lb_1] }\right)^2} \\
&+ { \dotlb[\lb_1, \lb_3]^2  \dotl[\la_2, \la_4]^6 \over (E^{13s})^2 (E^{24,s})^2}
\left(\dotlb[\lb_3, \lb_{2}] + i E^{24s} {\dotlm[\lb_3, \la_4] \over  \dotl[\la_4, \la_2]} \right)^2 \left(\dotlb[\lb_1, \lb_4]  + i E^{24s} {\dotlm[\lb_1, \la_2] \over \dotl[\la_2, \la_4]} \right)^2 \Bigg].  
\end{split}
\ee

The contribution from this partition can be written as
\be
\label{tu}
T_{u} = T_{u_1} + T_{u_2} + T_{u_3},
\ee
summing the contributions of the three terms on the right hand side, which 
are computed above.

\subsection{Final Answer}
The final answer for the stress tensor correlator can now be written
in terms of all the contributions above:
\be
\label{fullmhvstress}
T^{+-+-}(\vect{k_1}, \ldots \vect{k_4}) = T_s + T_t + T_u,
\ee
where the contributions from the three partitions are given in 
\eqref{ts}, \eqref{tt}, and \eqref{tu}. The formula above
involves only four structurally distinct expressions. These are the expressions for the residue at $w_{s_1}^+$, at $w_{s_3}$, at $w_{u_1}^+$ and
$w_{u_3}$. The formula above tells us that all other expressions are given by simple interchanges in these expressions.

\subsection{Flat Space Limit \label{subsecflatspacegrav}}
Although the final formulas above are more complicated than the corresponding formulas for currents, it is not difficult to extract the MHV graviton
amplitude from them. Our analysis is very similar to subsection \ref{subsecflatspacecurr}, so we will be brief here. 

First let us recall some facts about the flat space MHV graviton amplitude.
Consider the rational functions of $w$:
\be
\begin{split}
{\cal M}^{\text{MHV}}(w) &= i {\dotl[\la_4, \la_2]^6 \dotl[\lb_1, \lb_3] \over  \dotl[\la_3(w), \la_4] \dotl[\la_1(w), \la_2] \dotl[\la_2, \la_3(w)] \dotl[\la_1(w), \la_4] \dotl[\la_1(w), \la_3(w)]}, \\
{\cal M}^{\overline{\text{MHV}}}(w) &= i {\dotl[\lb_3, \lb_1]^6 \dotl[\la_2, \la_4] \over  \dotl[\lb_3,\lb_4(w)] \dotl[\lb_1,\lb_2(w)] \dotl[\lb_1, \lb_4(w)] \dotl[\lb_2(w),\lb_3] \dotl[\lb_2(w), \lb_4(w)]},
\end{split}
\ee
with $w$ extended according to \eqref{risagerextension}. The usual MHV graviton amplitude is given by $M^{\text{MHV}}(0) = M^{\overline{\text{MHV}}}(0)$ \cite{Berends:1988zp,Mason:2008jy,Nguyen:2009jk}.

The trick is to break $M^{\text{MHV}}(0)$ into partial fractions. This can be achieved by using the Cauchy theorem and writing $M^{\text{MHV}}(0)$ as the sum of the residues of $M^{\text{MHV}}(w)$ at its poles. However, 
these poles are precisely at the values of $w$ considered in our previous subsection. In particular, we have
\be
\label{mgravmhvpartial}
\begin{split}
M^{\text{MHV}}(0) &= \Bigg[\lim_{w \rightarrow w_{s_1}^+} {\dotl[\la_1(w), \la_2] \over \dotl[\la_1, \la_2]}  + \lim_{w \rightarrow w_{t_1}^+} {\dotl[\la_1(w), \la_4] \over \dotl[\la_1, \la_4]}  +  \lim_{w \rightarrow w_{s_2}^+} {\dotl[\la_3(w), \la_4] \over \dotl[\la_3, \la_4]}  \\ &+ \lim_{w \rightarrow w_{t_2}^+} {\dotl[\la_3(w), \la_2] \over \dotl[\la_3, \la_2]}  + \sum_{\pm} {w_{u_1}^{\mp} \over w_{u_1}^{\mp} - w_{u_1}^{\pm}} \lim_{w \rightarrow w_{u_1}^{\pm}} {\dotl[\la_3(w), \la_1(w)] \over \dotl[\la_3, \la_1]} \Bigg] M^{\text{MHV}}(w),
\end{split}
\ee
which is just the sum of the residues of $M^{\text{MHV}}(w)$ at all its poles. We can write down an analogous expression for $M^{\overline{\text{MHV}}}(w)$. 

Now, let us turn to our computations of the stress tensor correlator. To check the flat space limit, we need to focus on the coefficient of $(E^T)^{-3}$
and check that it matches the expressions above. However, a term of this kind
only comes from the ${1 \over E^T}$ term in  the derivatives. We can see a term of this form in \eqref{dsp} and \eqref{du1}. So, to check the coefficient we must only look at the {\em values of the integrands} (with the poles stripped off) at $E^T = 0$. 

This is easy to do. For example, consider \eqref{tpi1}. Near $E^T = 0$, we can see that 
\be
{\cal V}_{s_1}^+ = {\norm{k_1} \norm{k_2} \norm{k_3} \norm{k_4} \over 2 (E^T)^2} {\dotl[\lb_1, \lb_2(w_{s_1}^+)] \dotl[\la_4, \la_2]^6 \over \dotl[\la_1(w_{s_1}^+), \la_3(w_{s_1}^{+})]^2 \dotl[\la_4, \la_3(w_{s_1}^{+})]^2 \dotl[\la_1, \la_2]} + \ldots 
\ee
where the $\ldots$ are terms less singular in $E^T$. Here, we have just dropped the $E_{p}^{12}$ terms in \eqref{tpi1} and simplified other factors using $E^T = 0$.  Now, using the fact that $\la(w_{s_1}^{+}) \propto \la_2$, we can write
\be
{\cal V}_{s_1}^+
=   {\norm{k_1} \norm{k_2} \norm{k_3} \norm{k_4} \dotl[\lb_1, \lb_2(w_{s_1}^+)] \dotl[\la_4, \la_2]^6 \dotl[\la_2, \la_4] \over 2 (E^T)^2 \dotl[\la_1(w_{s_1}^+), \la_3(w_{s_1}^{+})] \dotl[\la_2, \la_3(w_{s_1}^{+})] \dotl[\la_1(w_{s_1}^{+}), \la_4] \dotl[\la_4, \la_3(w_{s_1}^{+})]^2 \dotl[\la_1, \la_2]} + \ldots 
\ee
Dropping terms that are less singular at $E^T = 0$, this becomes:
\be
{\cal V}_{s_1}^+= {1 \over 2 (E^T)^2} {\dotl[\lb_1, \lb_3] \dotl[\la_4, \la_2]^6  \over \dotl[\la_1(w_{s_1}^+), \la_3(w_{s_1}^{+})] \dotl[\la_2, \la_3(w_{s_1}^{+})] \dotl[\la_1(w_{s_1}^{+}), \la_4] \dotl[ \la_3(w_{s_1}^{+}), \la_4] \dotl[\la_1, \la_2]} + \ldots
\ee

Combining this with the ${2 i \over E^T}$ from ${\cal D}_{s_1}^+$ in \eqref{dsp}, we see
that 
\be
T_{s_1} =  {i  \over  (E^T)^3} {\norm{k_1} \norm{k_2} \norm{k_3} \norm{k_4} \dotl[\lb_1, \lb_3] \dotl[\la_4, \la_2]^6  \over \dotl[\la_1(w_{s_1}^+), \la_3(w_{s_1}^{+})] \dotl[\la_2, \la_3(w_{s_1}^{+})] \dotl[\la_1(w_{s_1}^{+}), \la_4] \dotl[ \la_3(w_{s_1}^{+}), \la_4] \dotl[\la_1, \la_2]} + \ldots
\ee
However, this is exactly the first term in \eqref{mgravmhvpartial}! Working through the other terms we see that the full gravitational correlator can be written
\be
T^{+-+-}(\vect{k_1},\ldots \vect{k_4}) = {\norm{k_1} \norm{k_2} \norm{k_3} \norm{k_4} \over (E^T)^3} \left(M^{\text{MHV}}(0) + M^{\overline{\text{MHV}}}(0)\right) + \ldots
\ee
where $\ldots$ are terms less singular in $E^T$. This is {\em precisely}
consistent with the flat space limit conjectured in \cite{Raju:2012zr}, which we remind the reader was: 
\be
M^{+-+-}(\vect{k_1}, \ldots, \vect{k_4}) = \lim_{E^T \rightarrow 0} {(E^T)^3 \over 2 \prod_{m=1}^4 \norm{k_m}} T^{+-+-}(\vect{k_1},\ldots \vect{k_4}).
\ee

This concludes our discussion of the flat space limit of the MHV graviton correlator. It is not difficult to see that for NMHV and N$^2$MHV configurations, the graviton correlator (with a pure Hilbert action) has no ${1 \over (E^T)^3}$ singularities and so its flat space limit vanishes just as one would expect.

\section{Discussion}
In this paper, we have obtained four point functions of the stress-tensor
in AdS$_4$/CFT$_3$ from a bulk gravity computation. Although the evaluation
of Witten diagrams in AdS is very complicated, we utilized the technique
devised in a companion paper to directly obtain the final answer 
using three point functions as an input. To our knowledge, this is the first time explicit expressions for the four point function of the stress-tensor have been written down. 

To summarize briefly, we wrote down a general formula for the four point
function for arbitrary external helicities and momenta in terms of the 
residues of a rational integrand at pre-specified poles. This formula 
is given in \eqref{formulafinal}. In the case of the MHV current and 
stress tensor correlators, we evaluated this formula in terms of the 
external spinors. The answer for the color-ordered MHV correlator is given
 in \eqref{colororderedmhv} and for the full MHV correlator is given 
in \eqref{fullmhvcurr}. The answer for the full MHV stress tensor correlator
is given in \eqref{fullmhvstress}. We also verified that, in the flat space limit, these answers give exactly the flat space MHV gluon and graviton
amplitudes. 

From a structural perspective, it is interesting that our recursion relations also remove the divergences
that usually appear in momentum space computations from the region near 
the boundary. What is striking, and related to this, is that our final
answers are purely rational functions of the momenta, their norms, and the norms of the sum of the momenta i.e. $\vect{k_m}, \norm{k_m}$ and $\norm{k_{m_1} + k_{m_2}}$. In particular, our answers are free of logarithms in momentum space.

The fact that stress tensor correlators have such an
analytic structure is also supported by the observation that their correlators
in AdS$_4$ can be obtained by doing a flat space computation (on half of flat space) using
conformal gravity \cite{Maldacena:2011mk}. Similarly, current correlators
in AdS$_4$ with pure Yang-Mills in the bulk can also be obtained by doing a computation on 4-dimensional flat space, cut off at $z = 0$. These computations do not lead to any logarithms in momentum space. 

Now logarithms in position space come from the fact that when we expand a four point correlator of some operator $\phi$ in a large-$N$ theory, $\langle \phi(0) \phi(\vect{x}) \phi(\vect{y_1}) \phi(\vect{y_2}) \rangle$, in terms of the contribution of various operators in the OPE when $x$ is close to $0$, we get terms like $\norm{x}^{\Delta_0 - 2 \Delta_{\phi} + \delta}$ where $\Delta_{\phi}$ is the dimension of $\phi$, $\Delta_O$ is the dimension of the operator
in the OPE-channel under consideration, and $\delta$ is a small ``anomalous dimension'' proportional to a negative power of $N$. Expanding this term in a ${1 \over N}$ expansion we get logarithms. From the fact that the Fourier transform of  $\norm{x}^{\Delta_{O} - 2 \Delta_{\phi} + \delta}$ is proportional to $\norm{k}^{-d - \Delta_{O} + 2\Delta_{\phi} - \delta}$, 
one might naively suspect that the absence of logarithms in momentum space is indicative of the absence of anomalous dimensions for double trace operators of the stress tensor. 

However, this logic is not quite correct.\footnote{I would like to thank Liam Fitzpatrick, Jared Kaplan and Joao Penedones for a discussion on this question.} In the case where ${\Delta_O - 2 \Delta_{\phi}}$ is a positive {\em even integer}, if we carefully consider this Fourier transform we find that we can get logarithms in position space without corresponding logarithms in momentum space. 
This is related to the fact that the Fourier transform of ${1 \over \norm{k}^{2 m + 3}}$ in $3$ dimensions, where $m$ is a non-negative integer, is proportional to $\norm{x}^{2 m} \log\left(\norm{x} \right)$ plus some terms that are analytic in $\vect{x}$ and depend on the precise $i \epsilon$ prescription that we use while performing the Fourier transform. In fact, the double trace operators of the stress tensor that appear in the four point correlator, which we have computed, give a contribution that is of
this form. Their contribution in the OPE can be written in the form
 $Q(\vect{x}, \vect{y_1}, \vect{y_2}) \norm{x}^{2 m + \delta}$ where $Q$ is a polynomial in $\vect{x}$. Consequently, they are subject to the subtlety above. It would be nice to perform this Fourier transform in full detail and extract the anomalous dimensions and OPE coefficients of double trace operators from our results.

As we mentioned above, the OPE provides another check on our final answer. The residue from the third pole in the list \eqref{polelist} automatically contains the product of three point functions of the stress tensor  multiplied with the appropriate power of the two point function. This accounts for the entire contribution of the conformal block of the stress tensor itself.\footnote{As we mentioned above, an advantage of momentum space is that we do not have to worry about conformal blocks: the contribution of all descendants in a channel is just obtained by multiplying correlators, on the left and the right, involving the primary.} It would be nice to show explicitly that the rest of the correlator is consistent with
the expectation that it comes from the contribution of double trace operators. 

Turning now to finer details in the four point correlator, we notice
that our answers for current correlators are relatively simple but our
expressions for stress-tensor correlators are still somewhat unwieldy. It would be nice to put these answers in their ``simplest possible'' form. One complication is that in momentum space it is only the Lorentz subgroup $SO(2,1)$,  of the 
conformal group on the boundary, that is manifest.  For scattering amplitudes in four dimensional flat space, the Lorentz group $SO(3,1)$ and a knowledge of how amplitudes scale under dilatations at tree-level is sufficient to ensure that four point function essentially depends only on one variable --- the scattering angle. This simplification cannot be obtained here just by using the Lorentz group of the boundary.

Of course the full conformal group is far more powerful but the constraints of special conformal invariance are differential equations in momentum space. This makes these constraints rather tedious to implement. 

One could also ask whether a knowledge of the flat space limit of the correlator, and some further assumptions about its analytic structure are enough to determine it completely. This deserves further attention. It is also possible that the simplest form of the four point correlators will be obtained by using another formalism, such as twistor space or a simpler set of recursion relations. We hope that the results presented here will be useful as inputs and checks on any such new method. So we have included a Mathematica program with the source of the arXiv version of this paper that allows for the automated evaluation of the formulas in this paper. 

We should mention that it is quite easy to generalize this formalism to supersymmetric theories.  While graviton tree-amplitudes are the same in supergravity and pure gravity, it is possible that expressing the amplitude in a manifestly supersymmetric form will make it more compact and also reveal links between correlators for different possible external helicities. We leave these investigations to future work. 

Finally, we should point out that in this paper we have observed several advantages of going to momentum space in AdS$_4$ for computations involving massless fields with spin. The wave-functions are very simple, and so $z$-integrals are easy to do; complicated interactions can be analyzed
by generalizing flat space techniques in momentum space; conformal blocks are trivial and it is also easy to take a flat space limit. This suggests that it would be very useful to analyze the Vasiliev theory in AdS$_4$ --- which contain massless higher spin fields ---  in momentum space. We leave this to future work. 

\section*{Acknowledgments}
I am grateful to Juan Maldacena and Guilherme Pimentel for collaboration in the early stages of this work. I would like to thank Sayantani Bhattacharyya, Bobby Ezhuthachan, Rajesh Gopakumar, Shiraz Minwalla, Kyriakos Papadodimas, Sandip Trivedi and Ashoke Sen for useful discussions. This work was primarily supported by a Ramanujan Fellowship of the Department of Science and Technology. I would also like to acknowledge the support of the Harvard University Department of Physics. I would like to thank the Institute for Advanced Study (Princeton), the Institute of Mathematical Sciences (Chennai), the Chennai Mathematical Institute, Delhi University and the Tata Institute of Fundamental Research (Mumbai) for their hospitality while this work was being completed.
\bibliographystyle{JHEP}
\bibliography{references}
\end{document}